\documentclass[twocolumn]{aastex701}
\usepackage{xspace}
\usepackage{amsmath}

\newcommand{\units}[1]{\ensuremath{\mathrm{{#1}}}}
\newcommand{\CaII}{\ion{Ca}{2}\xspace}
\newcommand{\Mbh}{\ensuremath{M_{\bullet}}\xspace}
\newcommand{\kms}{\units{km\,s^{-1}}\xspace}
\newcommand{\Msun}{\units{M_{\scriptstyle{\odot}}}\xspace}
\newcommand{\Lsun}{\units{L_{\scriptstyle{\odot}}}\xspace}
\newcommand{\Lsunv}{\units{L_{\scriptstyle{\odot},V}}\xspace}
\newcommand{\Mpc}{\units{Mpc}\xspace}
\newcommand{\kpc}{\units{kpc}\xspace}
\newcommand{\Angstrom}{\units{\text{\AA}}\xspace}
\newcommand{\MsunLsun}{\units{M_{\scriptstyle{\odot}}\,L_{\scriptstyle{\odot},V}^{-1}}\xspace}
\newcommand{\simsym}{\mathord{\sim}}
\newcommand{\approxsym}{\mathord{\approx}}

\begin{document}

\title{A Supermassive Black Hole Mass Measurement in NGC~5102 with Schwarzschild Orbit-superposition Modeling}

\correspondingauthor{Thomas K. Waters}
\email{waterstk@umich.edu}

\author[orcid=0000-0002-5231-7240]{Thomas K. Waters}
\affiliation{University of Michigan $\vert$ Department of Astronomy\\
1085 S University Ave, \\
Ann Arbor, MI 48109, USA}
\email{waterstk@umich.edu}

\author[orcid=0000-0002-1146-0198]{Kayhan Gültekin}
\affiliation{University of Michigan $\vert$ Department of Astronomy\\
1085 S University Ave, \\
Ann Arbor, MI 48109, USA}
\email{kayhan@umich.edu}

\author[orcid=0000-0002-8433-8185]{Karl Gebhardt}
\affiliation{University of Texas at Austin $\vert$ Department of Astronomy\\ 
2515 Speedway, \\
Austin, TX 78712, USA}
\email{gebhardt@astro.as.utexas.edu}

\author[orcid=0009-0002-3845-8175]{Soch Foskic}
\affiliation{University of Michigan $\vert$ Department of Astronomy\\
1085 S University Ave, \\
Ann Arbor, MI 48109, USA}
\email{sfoskic@umich.edu}

\author[orcid=0009-0000-1797-4950]{Ahmad Kadri}
\affiliation{University of Michigan $\vert$ Department of Astronomy\\
1085 S University Ave, \\
Ann Arbor, MI 48109, USA}
\email{amdkadri@umich.edu}

\shorttitle{Stellar-Dynamical SMBH Mass Measurement in NGC~5102}
\shortauthors{Waters et al.}


\received{May 18, 2026}
\revised{June 29, 2026}
\accepted{July 3, 2026}

\begin{abstract}
    We present a stellar-dynamical mass measurement of the central black hole in the lenticular galaxy NGC~5102 (SA0$^-$). Our analysis combines high-quality integral-field spectroscopy from the VLT Multi Unit Spectroscopic Explorer with high-spatial- and high-spectral-resolution \textit{HST} Space Telescope Imaging Spectrograph observations, using the \ion{Ca}{2} triplet as a stellar kinematic tracer. We constrain the black hole mass with axisymmetric, three-integral Schwarzschild orbit-superposition models, incorporating surface brightness measurements from \textit{HST} F547M WFPC2 imaging. Assuming a distance of $3.66\,\mathrm{Mpc}$, we find a black hole mass of $(1.30^{+0.19}_{-0.18})\times10^6\,\mathrm{M_{\scriptscriptstyle\odot}}$, which is within $1.7\sigma$ of a previous CO band-head-based Jeans Anisotropic Modeling result ($M_{\bullet}=(9.1^{+1.8}_{-1.5})\times10^5\,\mathrm{M_{\scriptscriptstyle\odot}}$). Our measurement is also consistent with literature extrapolations of the $M_{\bullet}$--$\sigma_e$ relation into the currently under-sampled low-mass regime. The close agreement between these independent dynamical approaches provides external validation of the Jeans Anisotropic Modeling framework and supports the robustness of our Schwarzschild orbit-superposition result, bolstering confidence in future black hole mass measurements with this framework.
    
\end{abstract}


\section{Introduction} \label{sec:intro}

\begin{figure*}
    \centering
    \includegraphics[width=0.8\paperwidth]{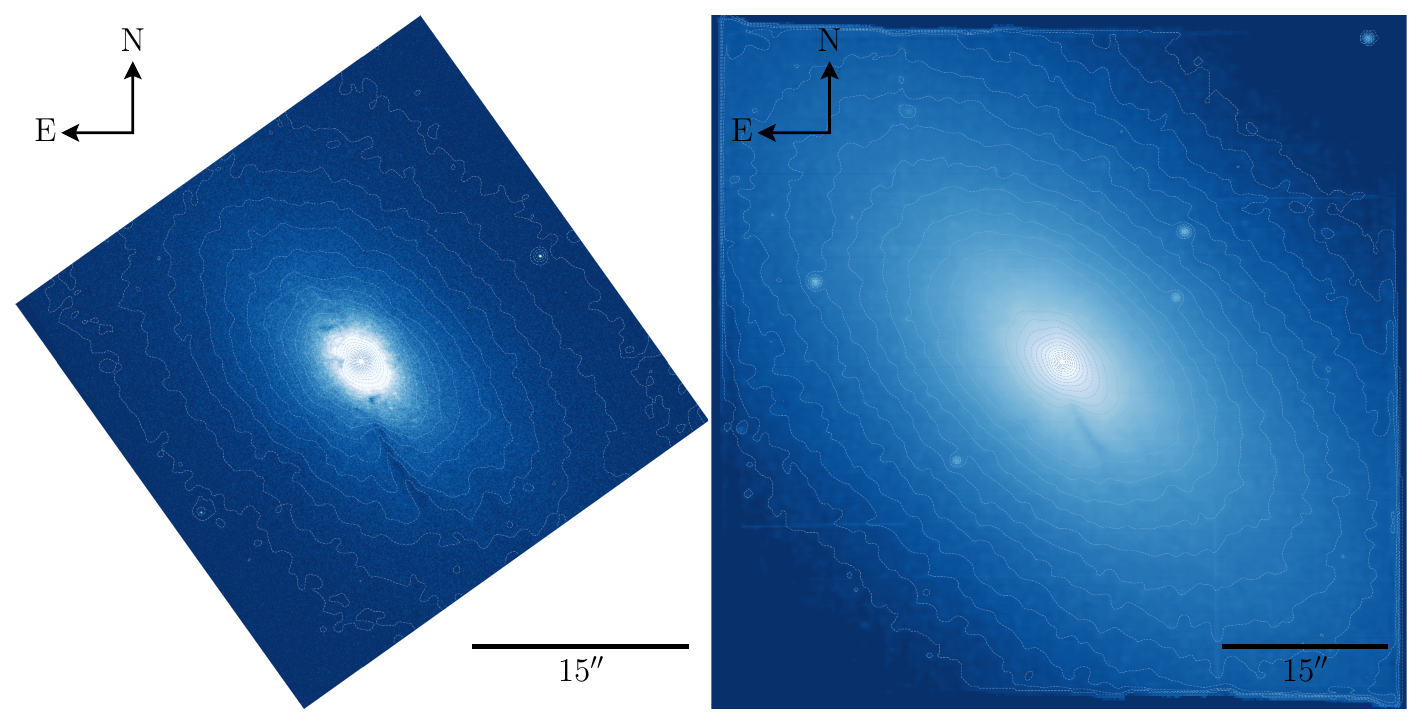}
    \caption{Image of NGC~5102 from \textit{HST}/WFPC2 in the F547M filter (left) and an optical white-light image from VLT/MUSE in the WFM (right). Both images have surface brightness contours (dotted lines) overplotted. The \textit{HST}/WFPC2 image, with a FoV of $28\farcs5\times28\farcs5$ $\left(505\times505\,\mathrm{pc}^2\right)$, is composed of four 20 second exposures and four 120 second exposures that were Nyquist resampled to a pixel scale of $0\farcs023$/pixel and combined with a sigma-clipped mean (see Section~\ref{sec:hst_data} for details). The MUSE data cube, shown here as a white-light image (summed along the wavelength axis from $4750\,\Angstrom$ to $9340\,\Angstrom$), is composed of four individual exposures with a total exposure time of 3840 seconds that were combined as part of the MUSE-DEEP program and covers a FoV of $1 \arcmin\times1\arcmin$ $\left(1.1\times1.1\,\kpc^2\right)$. Both images show several foreground stars and strong absorption by a dust lane approaching the galactic center. Smaller dust clouds near the center are visible in the higher-resolution \textit{HST} image.}
    \label{fig:images}
\end{figure*}   

    Supermassive black holes (SMBHs; $\Mbh \gtrsim 10^6\,\Msun$) have been shown to be ubiquitous in massive galaxies \citep{Kormendy&Ho2013, Kormendy&Gebhardt2001, Magorrian1998, Richstone1998}. Because this work focuses on the low-mass end of the nuclear black hole population, we use the broader term massive black hole (MBH) to refer collectively to central black holes in the intermediate-mass and supermassive regimes. MBHs are thought to play an important role in galaxy evolution, both because their masses correlate with a wide range of host-galaxy properties and because accretion onto MBHs can provide a source of energetic feedback \citep[e.g.,][]{Kormendy&Ho2013, Fabian2012}. Low-mass galaxies are particularly important laboratories for studying MBH demographics, since their black hole occupation fractions and masses provide key constraints on black hole seeding models and on the low-mass extensions of MBH--galaxy scaling relations \citep[e.g.,][]{Greene2020, Greene2016, Reines2015}. The presence of MBHs in lower-mass, local galaxies, however, is far less well understood.
    
    Recent studies suggest that MBH occupation fractions drop sharply in lower-mass galaxies: for example, \citet{Gallo2019} used Chandra data for 326 early-type galaxies to constrain the local black hole occupation fraction ($f_{\mathrm{occ}}$) to be greater than $47\%$ for galaxies with $M_{\ast} \lesssim 10^{10}\,\Msun$ at the $1\sigma$ confidence level (with values below $27\%$ excluded at $5\sigma$). \citet{Zou2025} improved on these constraints with an expanded sample of 1606 galaxies within 50 $\Mpc$, also imaged with Chandra, finding near-complete occupation ($f_{\mathrm{occ}} > 93\%$ at $2\sigma$) in galaxies with masses of approximately $10^{11}$--$10^{12}\,\Msun$, but the occupation fraction rapidly declines to $66{\%}_{-7\%}^{+8\%}$ and $33{\%}_{-9\%}^{+13\%}$ in galaxies with masses of approximately $10^{9}$--$10^{10}\,\Msun$ and $10^{8}$--$10^{9}\,\Msun$, respectively. These constraints rely on assumptions about typical accretion rates of MBHs in low-mass galaxies, so the true occupation fraction may be substantially higher if MBHs are more quiescent than current models allow. 

\begin{figure*}
    \centering
    \includegraphics[width=0.8\paperwidth]{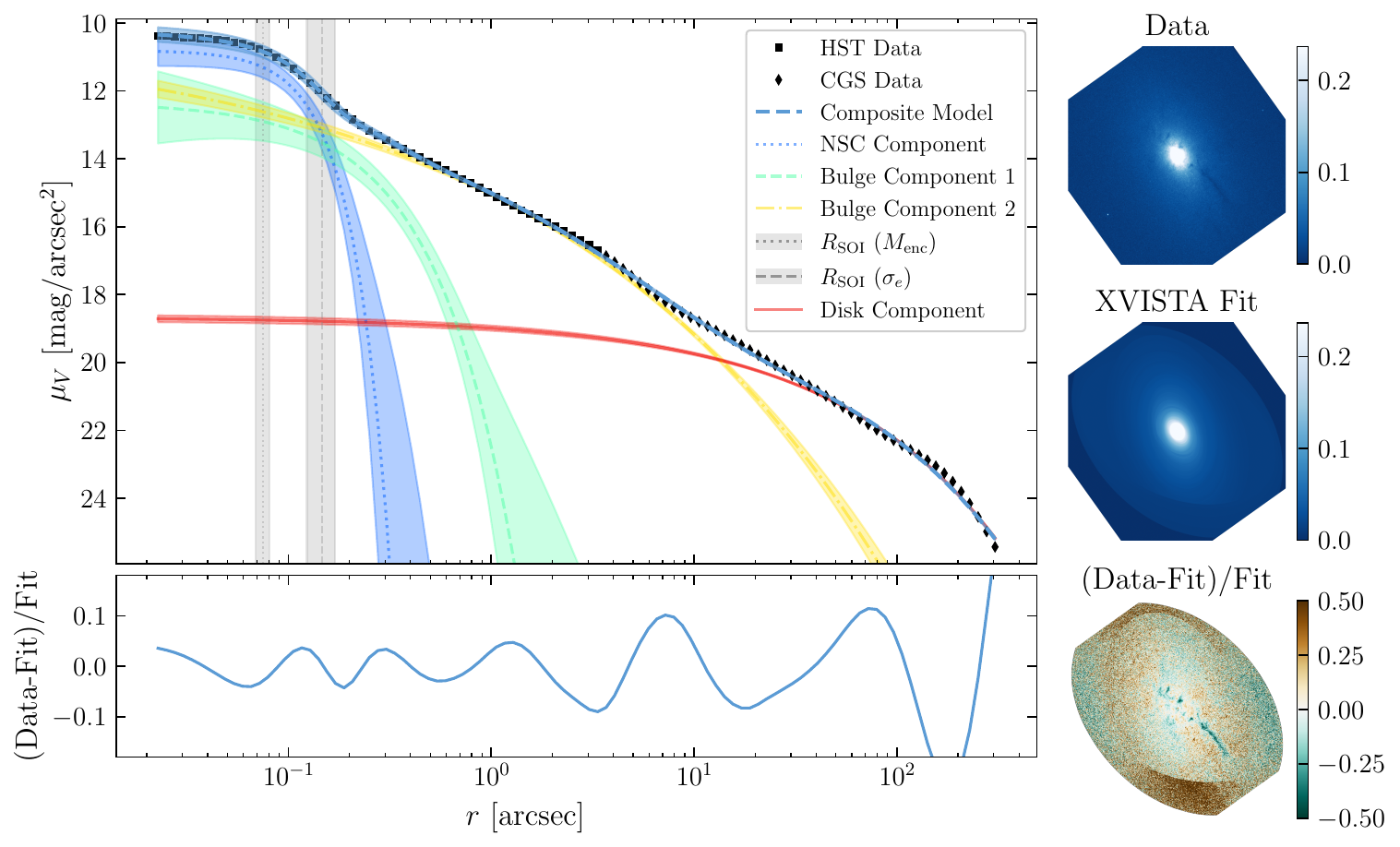}
    \caption{Summary of the stellar surface-brightness profile of NGC~5102 measured with the XVISTA software package (see Section~\ref{sec:surface_brightness}). The upper left panel shows the one-dimensional, azimuthally averaged stellar surface brightness profile measured from F547M \textit{HST} imaging data (black squares) concatenated with CGS data (black diamonds). Also shown are the best-fit Sérsic profile components fitted to the NSC (blue dotted line), the bulge (green dashed line and yellow dotted-dashed line), and the disk (red solid line) with their 16th and 84th percentile errors (filled bands). The sum of these components, the composite stellar surface brightness model, is shown as a cyan dashed line. Finally, the upper-left panel shows the spheres of influence for the central SMBH estimated with both the enclosed stellar mass method (gray dotted line) and the effective velocity dispersion method (gray dashed line) as well as their 16th and 84th percentile uncertainties. The bottom-left panel shows the residuals between the stellar surface-brightness profile and the composite Sérsic model. The right three panels show the \textit{HST} imaging data, the two-dimensional XVISTA surface brightness fit, and the fit residuals from top-right to bottom-right, respectively. The azimuthally averaged profile is matched extremely well (median absolute fractional residual 0.26\%). The two-dimensional residuals are larger (16\%) as they are dominated by the dust structures in the galaxy, which are visible in the bottom-right panel as clumps of negative values.}
    \label{fig:sb}
\end{figure*}   

    In galaxies where MBHs are present, they have relatively minimal direct gravitational impact: the region in which the MBH dominates the kinematic properties of the galaxy is very small. These ``spheres of influence,'' generally defined as the region within which the enclosed stellar mass equals the mass of the central MBH, are typically less than $\simsym1\,\kpc$, even for the most massive MBHs \citep{Ricci2017}. Yet, MBH masses ($\Mbh$) exhibit strong correlations with several global galaxy properties including, but not limited to, stellar velocity dispersions \citep[$\sigma$;][]{Ferrarese2000, Gebhardt2000, Gultekin2009b}, total stellar mass \citep[$M_\ast$;][]{McConnell&Ma2013, Reines2015}, bulge mass \citep[$M_\mathrm{bulge}$;][]{Haring&Rix2004, Kormendy1995, Marconi&Hunt2003, Saglia2016}, dark matter halo mass \citep[$M_{\mathrm{DM}}$;][but see also \citealp{Kormendy2011}]{Ferrarese2002, Volonteri2011, Voit2024}, galaxy and bulge luminosities \citep[$L$;][]{Beifiori2012, Dressler1989, Kormendy1993, Kormendy1995, Magorrian1998}, and X-ray luminosity \citep[$L_X$;][]{Gaspari2019}. Given these strong correlations between MBH mass and galaxy properties, MBHs may not only directly impact galaxy evolution, but also co-evolve with their host galaxies \citep[e.g.,][]{Hopkins2007, Kormendy&Ho2013, Schawinski2007}. However, these correlations alone do not identify the underlying causal mechanism or its direction: $\Mbh$ may regulate host structure via feedback, the host potential/assembly may regulate $\Mbh$ growth, or both may reflect shared dependence on additional variables. Several recent studies have applied various causal-discovery methods in an attempt to disentangle these possibilities \citep[e.g.,][]{Jin2024, Jin2025a, Jin2025b, Pasquato2023}. In particular, \citet{Davis2026} find evidence that the inferred direction between $\Mbh$ and stellar velocity dispersion is phase-dependent---favoring $\Mbh \rightarrow \sigma_e$ in star-forming systems and $\sigma_e \rightarrow \Mbh$ in quenched systems---which cautions against interpreting (or extrapolating) a single, morphology-agnostic scaling relation. In this context, AGN feedback provides a plausible physical route by which MBHs could influence host-galaxy evolution and reproduce the observed correlations \citep[e.g.,][]{Fabian2012, Matteo2008, Netzer2015, Silk&Rees1998}.

\begin{deluxetable}{lllll}
\tablecaption{Multi-component Sérsic Profile Fit Parameters
\label{tab:sersic_params}}
\tablehead{
    \colhead{Component}      &
    \colhead{$R_e$ (arcsec)} &
    \colhead{$R_e$ (pc)}     &
    \colhead{$n$}            &
    \colhead{$I_e$ ($\Lsunv\,\mathrm{pc}^{-2}$)}
}
\startdata
NSC     & $0.080 \pm 0.003$ & $1.43   \pm 0.05$ & $0.42 \pm 0.12$ & $(1.38 \pm 0.44)\times10^6$ \\
Bulge 1 & $0.191 \pm 0.084$ & $3.39   \pm 1.49$ & $0.90 \pm 1.1$  & $(1.33 \pm 1.55)\times10^5$ \\
Bulge 2 & $5.79  \pm 0.61$  & $102    \pm 11$   & $3.75 \pm 0.46$ & $(3.27 \pm 0.64)\times10^3$ \\
Disk    & $107   \pm 2$     & $1902.3 \pm 36.0$ & $1.84 \pm 0.07$ & $52.5  \pm 2.22$            \\
\enddata
\tablecomments{
Best-fit parameters for the four-component Sérsic profile decomposition
of the surface brightness profile of NGC~5102. $R_e$ is the effective radius, given in both arcseconds and parsecs (assuming a distance of $3.66\,\Mpc$), $n$ is the Sérsic index, and $I_e$ is the surface brightness in units of $\Lsunv\,\mathrm{pc}^{-2}$ at the effective radius.
}
\end{deluxetable}

    Testing these connections ultimately requires relating galaxy observables to the intrinsic parameters that characterize black holes: in the context of the no-hair theorem, black holes are fully characterized by their mass ($\Mbh$), spin ($a$), and charge ($Q$) \citep{Bekenstein1971b, Bekenstein1972a, Bekenstein1972b, Bekenstein1972c, Bekenstein1973, Misner1973, Ruffini1971a}. Astrophysical black holes are expected to be effectively neutral due to rapid charge screening by the surrounding plasma \citep[e.g.,][]{Bekenstein1971b}, leaving mass and spin as the relevant parameters. Spin can, in principle, strongly influence the radiative efficiency and the power available for jets and feedback \citep[e.g.,][]{Blandford&Znajek1977, Tchekhovskoy2011}. In practice, spin is typically inferred through model-dependent techniques, including relativistic reflection modeling, particularly of broadened Fe~K$\alpha$ emission, thermal continuum fitting, and methods based on jet or outflow energetics; these approaches depend on assumptions about accretion-flow structure, disk truncation near the innermost stable circular orbit (ISCO), coronal geometry, magnetic-field strength, and complex absorption \citep[e.g.,][]{Brenneman2006, ReynoldsNowak2003, Reynolds2014}. Empirical trends involving inferred spin have been explored in sizable AGN samples \citep[e.g.,][]{Daly2019, Daly2022}, but a broadly accepted spin--host-galaxy scaling relation analogous to the standard $\Mbh$--galaxy relations has not yet emerged. Irrespective of any spin--galaxy connection, a large sample of robust $\Mbh$ measurements remains critical for quantitatively assessing $\Mbh$--galaxy coevolution.

    A broad set of techniques is used to estimate MBH masses, spanning measurements that directly constrain the central gravitational potential and methods that infer $\Mbh$ from secondary observables. Dynamical approaches include stellar and gas kinematics, while reverberation mapping provides a complementary, variability-based route in active systems \citep[e.g.,][]{Macchetto1997, Peterson2004, Gebhardt2011}. Indirect estimators instead rely on empirical calibrations with quantities such as host scaling relations, X-ray scaling methods, or the fundamental plane \citep[e.g.,][]{Bentz2010, Merloni2003, Gliozzi2011, Gultekin2019, McConnell&Ma2013, Walsh2013, Woo2013}. Because these indirect approaches must be anchored to directly calibrated samples, expanding and homogenizing the set of dynamical measurements remains essential, especially where extrapolation is unavoidable. Stellar dynamics is particularly valuable in this context: it can be applied to quiescent nuclei and to galaxies lacking the cold gas or masing disks required for other direct techniques, making it one of the most broadly applicable routes to direct MBH masses \citep{vanderMarel1998, Gebhardt2000}.
    
\begin{figure*}
    \centering
    \includegraphics[width=0.8\paperwidth]{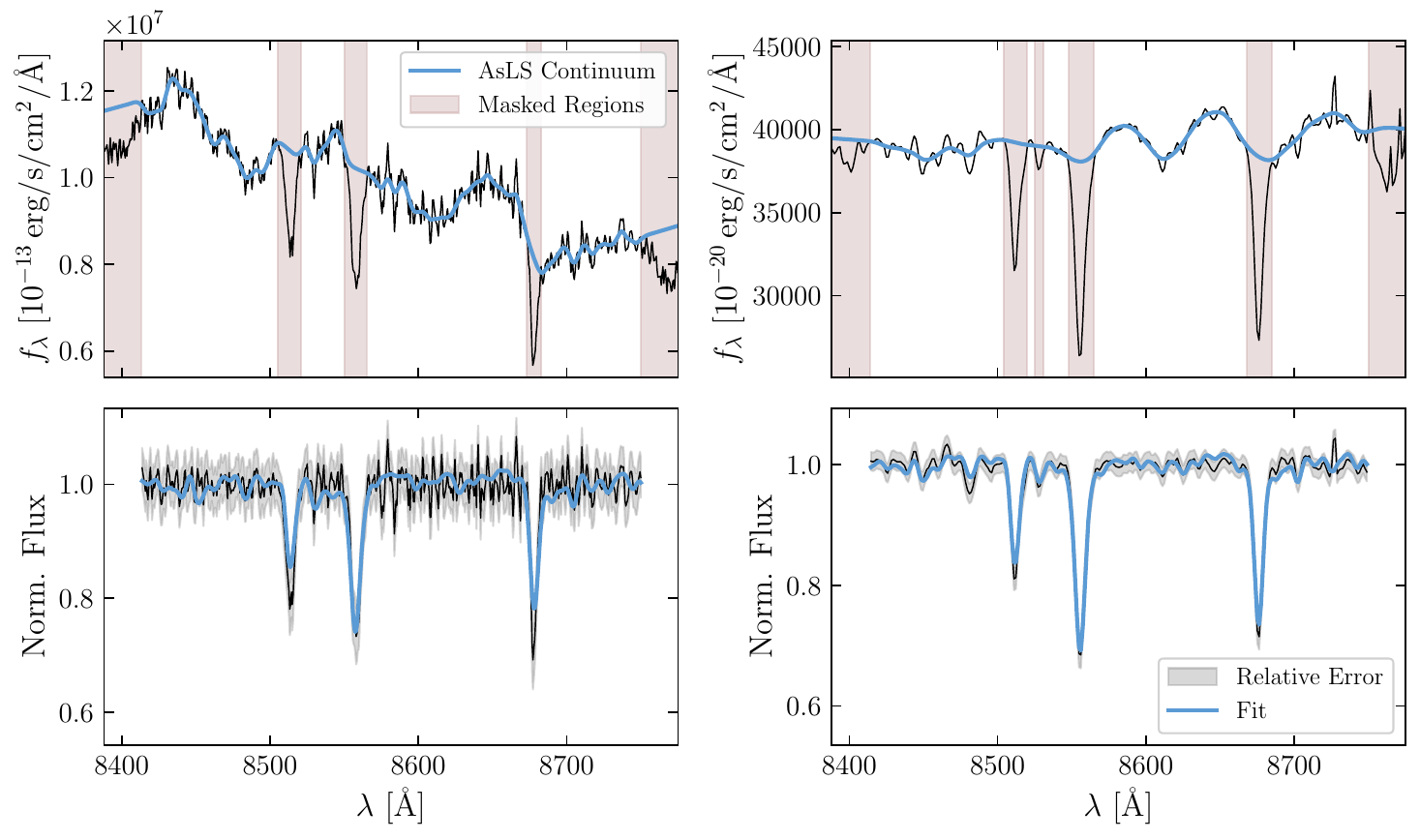}
    \caption{Spectra from the innermost bin for both the STIS data (left column) and MUSE data (right column). The top row shows the binned spectra in the \CaII\ triplet range ($8375$--$8775\,\Angstrom$) with an AsLS continuum fit (blue line). The masked regions (brown filled bands) were excluded from the continuum fit. The second row shows the continuum-normalized spectra overlaid with spectral fits (blue line) using the penalized-likelihood template-convolution procedure described in Section~\ref{sec:kinematic_extract}. The normalized spectra are plotted with their relative errors (gray filled bands) estimated from the RMS of the observed spectra added in quadrature to the uncertainties from the AsLS continuum fit. The spectral fits reproduce the overall continuum and the \CaII\ triplet features well. While the \CaII\ triplet depths are underestimated in some instances, the fits remain within the relative error of the normalized spectra and accurately capture the widths of the \CaII\ triplet lines, which directly constrain the velocity dispersions used as input to the SCO modeling (see Section~\ref{sec:sco_modeling}).}
    \label{fig:spectra}
\end{figure*}  

    Cross-checking MBH masses obtained with different observables and assumptions is one of the most effective ways to isolate systematic errors and to understand where particular estimators fail. For instance, \citet{Williams2023} found broad consistency among direct techniques across their comparison sample, yet highlighted cases (e.g., NGC~4151) where indirect estimates diverge by more than an order of magnitude. Using hard X-ray AGN measurements, \citet{Gliozzi2024} similarly showed that while some indirect approaches (e.g., fundamental-plane and X-ray scaling relations) can track dynamical masses, commonly used host-based relations may be biased in specific regimes; for example, $\Mbh$--$\sigma$ can over-predict and $\Mbh$--$M_\ast$ can under-predict in some AGN environments \citep{Reines2015}. These results motivate additional direct dynamical measurements in the low-mass regime, where the calibration sample is sparse and the combination of increased intrinsic scatter, morphology dependence, and selection effects can make extrapolations especially uncertain (see Section~\ref{sec:discussion}).
    
    These tensions become even more pronounced in the low-mass regime, where some scaling relations appear inconsistent with direct measurements. For example, \citet{Greene2016} found that black hole masses measured in maser galaxies can lie more than an order of magnitude below expectations from the extrapolated $\Mbh$--$M_\ast$ relation. In addition, galaxy--black hole scaling relations show increased scatter and potentially different overall behavior in the low-mass regime of central MBH masses. Examples include systematically under-massive MBHs in late-type galaxies relative to the $\Mbh$--$M_\mathrm{bulge}$ relation \citep{Greene2016, Lasker2016}, as well as evidence for a break in the $\Mbh$--$M_\mathrm{bulge}$ relation, where ``Sérsic'' galaxies, i.e., elliptical galaxies that are lacking central cores, and spiral galaxies that follow a steeper slope than massive early-type galaxies \citep{Savorgnan2016, Scott2013}. These inconsistencies highlight the need for additional direct black hole mass measurements, particularly in low-mass galaxies, to establish whether apparent departures reflect true population differences or method- and/or selection-dependent systematics.

    In addition to the need to accurately constrain the behavior of scaling relations across the MBH mass spectrum, the formation of MBHs themselves is another unanswered question. Current models of MBH formation generally invoke three broad ``seeding'' channels: Population III stellar remnants, direct collapse, and gravitational runaway. These seeding mechanisms have been explored extensively in models of MBH formation and evolution \citep[e.g.,][]{Loeb1994, Fryer2001, Haiman2001, Portegies2002, Bromm2003, Begelman2006, Lodato2006, Chantavat2023}, each of which yields different predictions for the BH mass function and occupation fraction in low-mass galaxies. Therefore, improving the demographics of low-mass BHs in galactic centers is critical for constraining the dominant mechanisms of SMBH seeding. Because of their low stellar masses, dwarf galaxies are particularly important systems for BH mass measurements to constrain seeding mechanisms \citep{Greene2020}.

\begin{figure*}
    \centering
    \includegraphics[width=0.8\paperwidth]{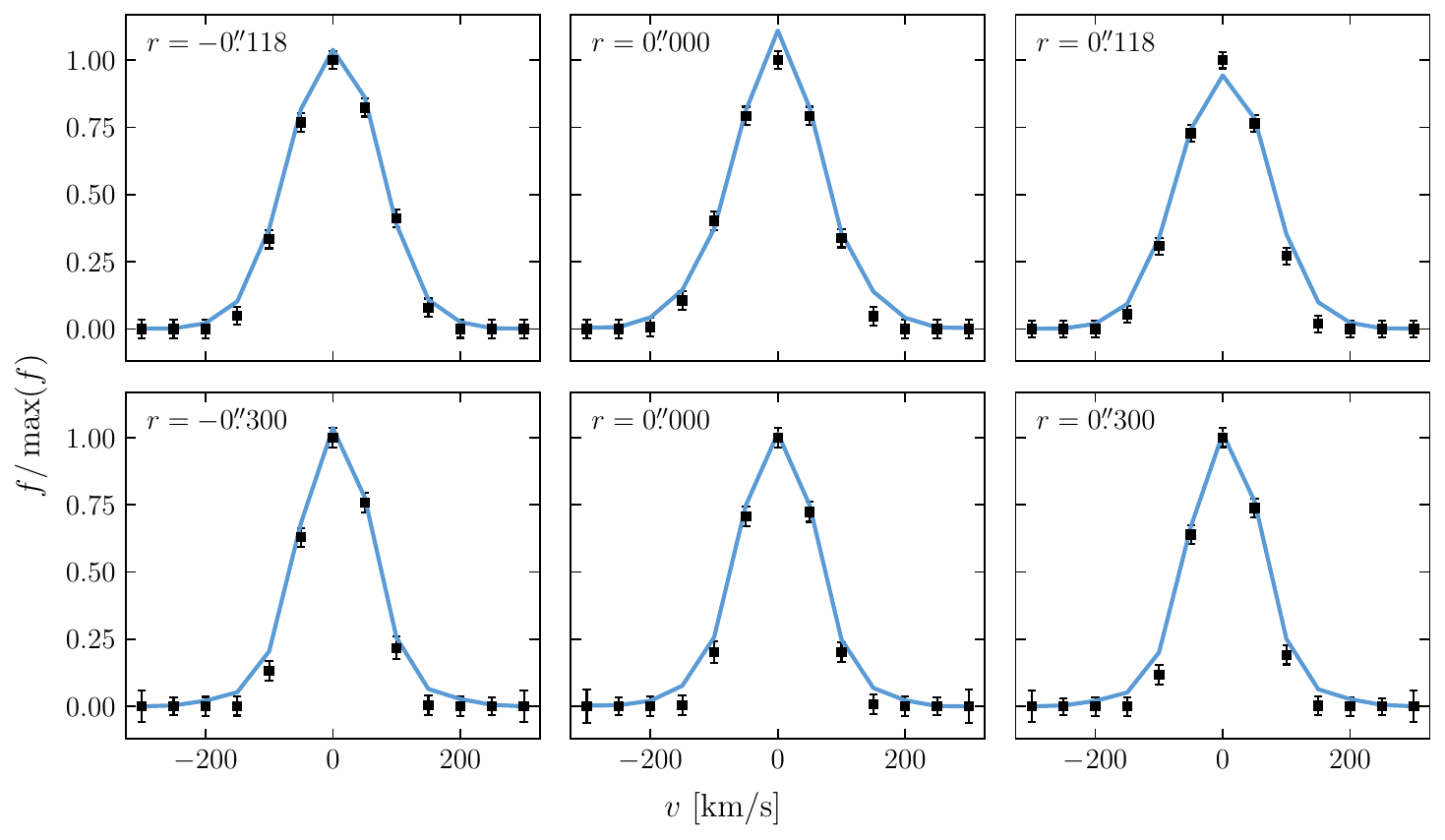}
    \caption{Six example LOSVDs ($f$; the fraction of line-of-sight stellar velocities between $v$ and $v+\Delta v$, where $\Delta v$ is $50\,\kms$) for the three innermost radial bins (as annotated on the plot) along the kinematic major axis extracted from the normalized STIS spectra (top row) and MUSE spectra (bottom row). The LOSVDs extracted from the data (black data points) are compared to those from our best-fit SCO model (blue lines). While there are minor deviations, particularly in the wings of the LOSVDs, the excellent overall fit quality ($\Delta \chi^2=90$ compared to the best-fit no-black-hole model) strongly supports the inference that a central black hole of mass $(1.30^{+0.19}_{-0.18})\times10^6\,\Msun$ is required to reproduce the observed kinematic data.}
    \label{fig:losvd}
\end{figure*}   

    Furthermore, to develop a full picture of MBH formation and evolution, it is crucial to explore black holes across the full range of masses. In this context, intermediate-mass black holes (IMBHs; here defined as $10^2\,\Msun<\Mbh<10^6\,\Msun$) occupy a potential ``missing link'' between stellar-mass black holes and SMBHs. IMBHs are frequently invoked in seeding scenarios as progenitors that can grow into SMBHs, helping to explain the existence of $\gtrsim10^9\,\Msun$ quasars within the first $\simsym700\,\mathrm{Myr}$ of cosmic time \citep{Banados2018}. While black holes have been reported up to $\simsym4\times10^{10}\,\Msun$ \citep{Ghisellini2009}, detections of IMBHs with high statistical confidence remain elusive. In addition to their relevance for seeding models, IMBHs may contribute to the gravitational-wave source population accessible to upcoming space-based instruments such as \textit{LISA}. Independent demographic constraints in this mass regime are therefore important for interpreting future gravitational-wave observations.

\begin{figure*}
    \centering
    \includegraphics[width=0.8\paperwidth]{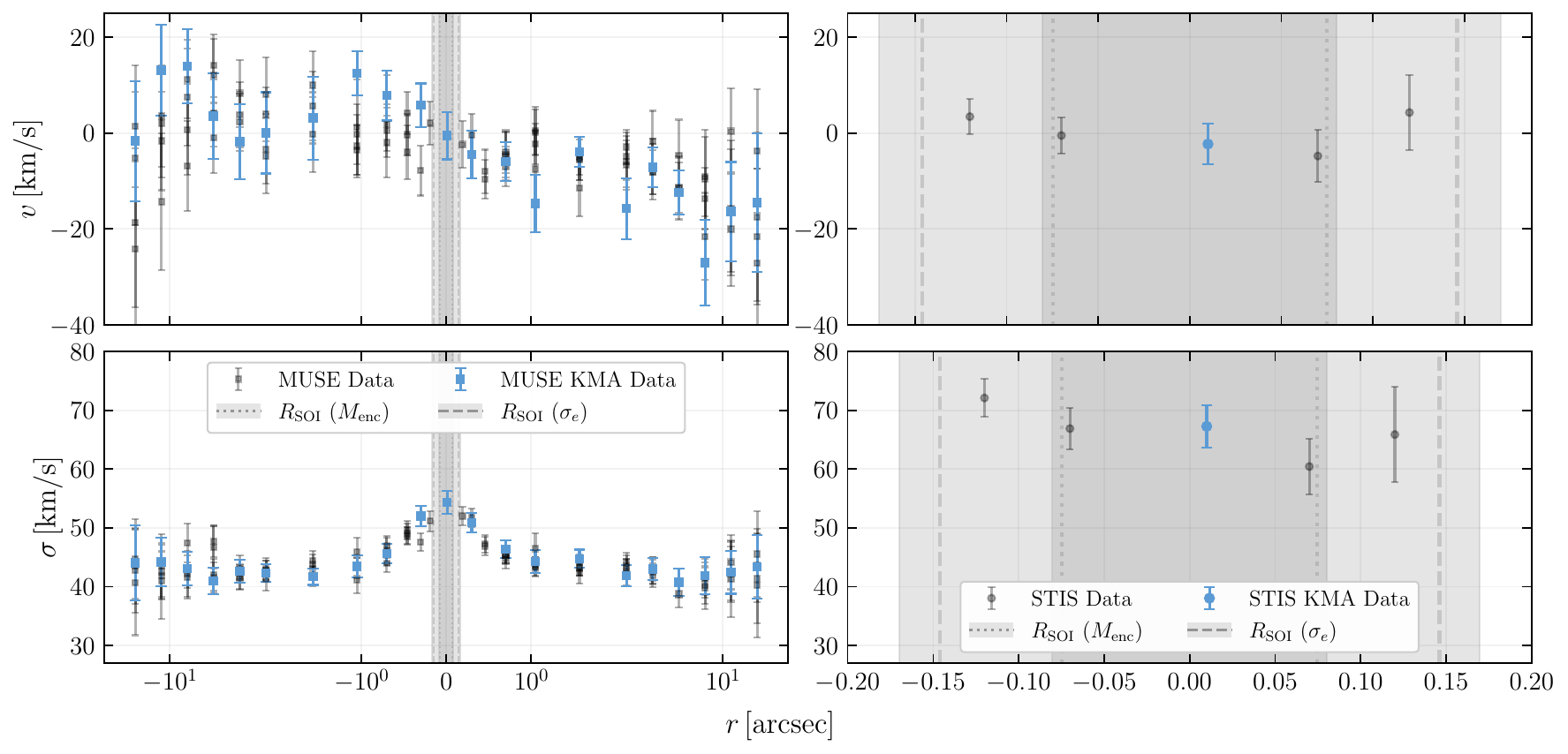}
    \caption{Velocity $v$ (top) and velocity dispersion $\sigma$ (bottom) profiles for MUSE (left, squares) and STIS (right, circles). Blue symbols show bins along the KMA; black points show data in the remaining angular bins. Gray shaded regions indicate $R_{\mathrm{SOI}}$ calculated from $M_{\mathrm{enc}}$ (dotted) and $\sigma_e$ (dashed). The velocity dispersions derived from STIS spectra within $R_{\mathrm{SOI}}$ reach $\simsym 70\,\kms$ in the innermost bin, consistent with the presence of a central black hole. The velocity dispersions from MUSE data peak at $\simsym 55\,\kms$ in the central bin, showing suppression relative to STIS. This difference arises from the significantly different pixel scales: MUSE has $0\farcs2$ pixels, larger than $R_{\mathrm{SOI}}$ ($0\farcs075 \pm 0\farcs006$ or $1.33\pm0.10$ pc) based on $M_{\mathrm{enc}}$, while STIS has $0\farcs050$ pixels, well within both estimates of $R_{\mathrm{SOI}}$. Therefore, MUSE averages over a larger physical region, and thus more stellar orbits, in the innermost pixel, suppressing the peak of the velocity dispersion profile.}
    \label{fig:kinematics}
\end{figure*}   

    In this work, we present a central MBH mass measurement for the lenticular galaxy (SA0$^-$) NGC~5102 using a three-integral Schwarzschild orbit-superposition modeling method (SCO; see Section~\ref{sec:sco_modeling}) with archival data from the \textit{Hubble Space Telescope}'s (\textit{HST}) Wide-Field and Planetary Camera 2 (WFPC2) and Space Telescope Imaging Spectrograph (STIS), together with data from the Very Large Telescope's (VLT) Multi Unit Spectroscopic Explorer (MUSE). Our work employs the surface-gravity-independent \CaII triplet as a tracer for stellar dynamics due to its strong spectral presence in K and G-type giant stars, which dominate the luminosity of galaxies, and its relative insensitivity to extinction \citep{Silge2003}. NGC~5102 has previously been the subject of a stellar-dynamical mass measurement using Jeans Anisotropic Modeling (JAM) based on CO band-head absorption features \citep{Nguyen2018, Nguyen2019}, which independently yielded $\Mbh=(9.1^{+1.8}_{-1.5})\times10^5\,\Msun$. We compare our results in detail in Section~\ref{sec:discussion}.
    
    This paper is organized as follows. Section~\ref{sec:observations} describes the observations and data reduction for the VLT/MUSE and \textit{HST} (WFPC2 and STIS) datasets. In Section~\ref{sec:surface_brightness}, we present the measurement and decomposition of the $V$-band stellar surface brightness profile using \textit{HST}/WFPC2 F547M imaging and its integration with the wide-field Carnegie Galaxy Survey (CGS) data. Section~\ref{sec:kinematic_extract} outlines our extraction of stellar kinematics from the \textit{HST}/STIS and VLT/MUSE spectra and our determination of the kinematic major-axis position angle. In Section~\ref{sec:modeling}, we describe our radially varying mass-to-light ratio ($M/L$) model and our SCO modeling framework, and we present the resulting constraints on the central black hole mass in NGC~5102. In Section~\ref{sec:discussion}, we examine the kinematic evidence that drives the preference for a nonzero $\Mbh$, discuss how the nuclear star cluster and the resulting $M/L$ gradient affect the mass inference and the interpretation of the sphere of influence, and compare our Schwarzschild-based measurement to the prior JAM-based result of \citet{Nguyen2018, Nguyen2019}. We then place NGC~5102 on the \citet{Kormendy&Ho2013} (hereafter KH13) $\Mbh$--$\sigma_e$ relation by computing $\sigma_e$ from our measured kinematic profiles, and we discuss how this low-$\sigma_e$ data point relates to the increased scatter and possible morphology- and selection-dependent trends suggested by existing dynamical samples. For NGC~5102, we assume a luminosity distance of $3.66\,\Mpc$, a weighted average of existing TRGB measurements \citep{Jacobs2009, Tully2016, Tully2023}.

\section{Observations and Data Reduction} \label{sec:observations}

    We analyzed archival observations of the lenticular galaxy NGC~5102 (SA0$^-$), obtained with MUSE on the ESO VLT, and with WFPC2 and STIS aboard \textit{HST}. Figure~\ref{fig:images} shows our reduced \textit{HST}/WFPC2 PC1 F547M image and a MUSE white-light image (made by summing the data cube along the wavelength axis). Overplotted on both images are surface brightness contours, highlighting the presence of a dust lane near the galactic center. The \textit{HST} image has been rotated to match the north-up, east-left orientation of the MUSE data cube on the sky.

\subsection{MUSE Observations and Data Reduction} \label{sec:muse_data}

    NGC~5102 was observed with MUSE on UT4-Yepun at the ESO VLT during the science verification run, under program 60.A-9308(A) (PI: Mitzkus). The wide-field mode delivers a $1\arcmin\times1\arcmin$ FoV at a plate scale of $0\farcs2$/pixel, spanning $4750$--$9340\,\Angstrom$ with $1.25\,\Angstrom$ fixed spectral sampling and $R \approxsym 3000$ (FWHM $\simsym2.5\,\Angstrom$) \citep{Bacon2010}. Four dithered and rotated exposures ($960$\,s each; $90^\circ$ rotations) were collected between 2014 June 23 and 24, totaling $3840$\,s on-target. The mean elevation of the exposures was $51^\circ$. Observing blocks alternated between NGC~5102 and $300$\,s offset sky exposures to enable precise sky subtraction given the galaxy's large apparent size compared to the FoV.
    
    Reduction and calibration of the MUSE observations were performed by the ESO Science Archive team using the official MUSE pipeline, as documented in the MUSE Data Reduction System \citep{Weilbacher2012, Weilbacher2020}. Processing included bias subtraction, flat-fielding, wavelength calibration, geometric distortion correction, sky subtraction (using offset exposures), and flux calibration with spectrophotometric standards per ESO protocols. The fully reduced, sky-subtracted, and flux-calibrated data cubes are available as public MUSE-DEEP Phase 3 data products on the ESO Science Archive \citep{MUSEDEEP2017, muse_eso_data}.
    
    To ensure robust error propagation throughout our modeling framework, we addressed a known MUSE pipeline limitation: variance arrays are underestimated due to the introduction of correlated errors during drizzle resampling from \texttt{PIXTABLE} format to data cubes. Following \citet{Bacon2017} and \citet{Sanderson2021}, we scaled the variance array of our cube by a factor $2.78$, an empirically derived correction for the drizzle $\texttt{pixfrac}=0.8$ configuration. Although this adjustment does not account for correlated errors, resampling tests yielded negligible differences in the variance of the final, binned spectra used in our kinematic extraction (Section~\ref{sec:kinematic_extract}).
        
\subsection{\textit{HST} Observations and Data Reduction} \label{sec:hst_data}

    Archival \textit{HST} observations of NGC~5102 were obtained under program 8591 (PI: Richstone) in 2001--2002, using both WFPC2 and STIS. These datasets provide high spatial and spectral resolution coverage of the central regions of NGC~5102. The WFPC2 and STIS observations used in this work are available from MAST at \dataset[doi:10.17909/1KDP-S611]{https://doi.org/10.17909/1KDP-S611} \citep{hst_data}.

\subsubsection{STIS Spectroscopy} \label{sec:hst_spec}

    \textit{HST}/STIS obtained two dithered long-slit spectra on 2002 January 13 using the G750M grating ($52\arcsec\times0\farcs1$ aperture), centered at $8561\,\Angstrom$, with spectral resolving power $R \approxsym 5000$. The $0\farcs1$ slit width yields a spatial scale of $0\farcs050$/pixel, and the spectral FWHM is approximately $1.1$--$1.6\,\Angstrom$ for extended sources. Two exposures ($1760$\,s and $2832$\,s) were acquired for a total integration time of $4592$\,s. The slit was dithered between exposures to mitigate detector artifacts \citep{Kimble1998, Woodgate1998}.

    Data reduction used the standard \textit{HST} \texttt{calstis} pipeline. Steps included bias subtraction, flat-fielding, wavelength and heliocentric calibration, and flux calibration with sensitivity functions. The reduced spectra from the two STIS exposures were aligned by matching the slit position with the highest total flux in each exposure, shifted onto a common spatial grid, and combined using exposure-time weighting. Residual outliers deviating by more than $3\sigma$ were removed with iterative sigma clipping.
    
\subsubsection{WFPC2 Imaging} \label{sec:hst_img}

    High-resolution imaging was obtained with WFPC2/PC1 \citep{Holtzman1995} in the F547M filter (central $\lambda$ $5483\,\Angstrom$, pixel scale $0\farcs046$/pixel). Eight exposures (four at $20$\,s, four at $120$\,s; total $560$\,s) were taken on 2001 May 27, following an \textit{HST} spiral dither pattern with $0.5$\,pixel shifts.

    WFPC2 data reduction employed the \textit{HST} calibration pipeline \citep{Gonzaga2012}. Bad pixels were identified from data quality (DQ) arrays in the image headers, which flag various instrumental defects including permanent dead pixels, hot pixels, cosmic-ray hits, and charge-bleeding artifacts. Flagged bad pixels were replaced via median filtering over a $5\times5$ pixel neighborhood, preserving flux while maintaining local image structure. Cosmic rays were detected by comparing each aligned image against the pixel-wise median of the entire stack; pixels deviating by more than $3\sigma$ from the median were identified as cosmic ray-affected and replaced with the median value. An exclusion region around the galaxy nucleus, with a radius of 10 pixels, was maintained during cosmic ray detection to preserve real astrophysical structure.
    
    Images were aligned using dither offsets recorded in the image headers, with sub-pixel shifts applied in Fourier space to preserve photometric accuracy. For photometric calibration, count rates (obtained by dividing raw counts by exposure time) were converted to absolute flux densities using WFPC2 calibration keywords and the STMAG magnitude system (see Section~\ref{sec:surface_brightness} for more detail). 
    
\begin{figure}
    \centering
    \includegraphics[width=0.39\paperwidth]{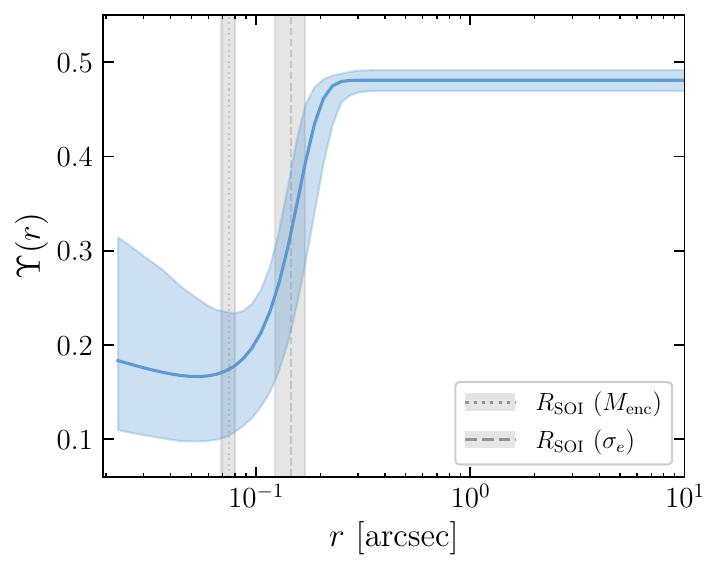}
    \caption{Radial profile of the stellar $M/L$ implied by our best-fit bulge $V$-band $M/L$ ($\Upsilon_V=0.48\pm0.01 \,\MsunLsun$) and NSC-to-bulge $M/L$ ratio ($\xi=0.025^{+0.021}_{-0.016}$). The blue line shows the median profile with $1\sigma$ uncertainty band (shaded region). Gray shaded regions indicate $R_{\mathrm{SOI}}$ estimated from $M_{\mathrm{enc}}$ (dotted) and $\sigma_e$ (dashed). The uncertainties were calculated by propagating the statistical uncertainties from the modeling and the uncertainties in the surface brightness profile (see Section~\ref{sec:surface_brightness}, Section~\ref{sec:vml}, and Figure~\ref{fig:sb}). This profile was derived to accurately model the changing $M/L$ induced by the central NSC, which is significantly bluer than the surrounding bulge, suggesting a recent burst of star formation in the galactic center. The presence of bright blue stars results in a sharp decrease in $\Upsilon(r)$ near the galactic center.}
    \label{fig:vml}
\end{figure}   

\subsubsection{Nyquist Sampling via Fourier-Space Upsampling} \label{sec:nyquist}

    To ensure Nyquist sampling of the point spread function (PSF) and achieve enhanced spatial resolution within the sphere of influence of the central massive black hole, the aligned images were upsampled by a factor of 2 in each dimension using Fourier-space interpolation with cubic spline kernels, yielding a final pixel scale of $0\farcs023$/pixel. This upsampling approach provides band-limited interpolation that prevents aliasing---critical for maintaining image fidelity.
    
    The upsampled images were processed in Fourier space through sub-pixel phase alignment using the Fourier shift theorem. Sub-pixel dither offsets (scaled to match the upsampled pixel scale) were encoded as phase ramps and applied to each exposure's fast Fourier transform (FFT), ensuring precise astrometric registration without spatial-domain resampling artifacts.
    
    Frequency-domain smoothing was applied using a smooth, circular, tapered filter in Fourier space:
\begin{equation}
    H(R)=\frac{1}{2}\left[1+\tanh\left(\frac{R_c-R}{w}\right)\right],
\end{equation}
    where $R=\sqrt{f_x^2+f_y^2}$ is the radial frequency coordinate, $R_c=0.4\times{f_{\text{Nyquist}}}$ is the frequency cutoff (40\% of Nyquist), and $w=0.05\times{f_{\text{Nyquist}}}$ is the taper width. This filter suppresses high-frequency noise above the effective resolution limit while the smooth $\tanh$ transition minimizes Gibbs ringing artifacts. The cutoff frequency corresponds to an effective spatial resolution of $\simsym$1.25 upsampled pixels, appropriate for the combined PSF of eight dithered WFPC2/PC1 observations.
    
    Inverse FFTs yielded reconstructed images. The final science image was created via a $3\sigma$-clipped mean combination of the eight reconstructed frames, providing robust cosmic ray rejection while preserving signal in low-surface brightness regions. The resulting image is critically sampled at the Nyquist scale, with a pixel scale of $0\farcs023$/pixel and effective spatial resolution $\simsym$1.25 pixels.
    
\section{Stellar Surface Brightness Profile} \label{sec:surface_brightness}

    We derived the surface brightness profile of NGC~5102 using XVISTA, a software package designed for the reduction and analysis of astronomical images and spectra \citep{XVISTA}. Among its many functions, XVISTA includes a robust framework for measuring the surface brightness profiles of astronomical images using iterative elliptical isophote fitting. The algorithm begins by sampling the image with circular annuli and computing the mean intensity of each contour. Along each contour, low-order sine and cosine transforms are taken to provide initial estimates of ellipticity, position angle, and isophotal center, which are then iteratively refined until the fitted ellipses converge smoothly and accurately trace the galaxy's observed isophotes. As it is imperative to accurately constrain the surface brightness in the innermost portion of the galaxy, XVISTA uses high-accuracy sinc interpolation to determine pixel values along the inner 15 isophotes. For the remainder of the galaxy, a less computationally intensive interpolation method is used. Once the isophotes are accurately fitted, XVISTA returns the average surface brightness along each isophote as well as the ellipticity and position angle of the elliptical fits. From these values, a two-dimensional surface brightness profile is then constructed.
    
    To extract the surface brightness profile from our mean-combined and upsampled \textit{HST} F547M image (see Section~\ref{sec:nyquist}), we corrected the image for geometric distortion within XVISTA using the WFPC2 polynomial distortion table. As is apparent in Figure~\ref{fig:images}, NGC~5102 displays strong absorption by a narrow dust lane and smaller dust clouds near the galactic center. To ensure accurate isophote fits, we masked these regions, as well as bright star clusters and foreground stars, prior to the elliptical isophote fitting. 

    The final surface brightness profile returned from XVISTA, in count-rate units, is then converted to physical flux density using WFPC2 calibration keywords: \texttt{PHOTFLAM}, the inverse sensitivity; \texttt{PHOTZPT}, the photometric zeropoint; and the STMAG system, where zero magnitude corresponds to a flux density of $3.63\times10^{-9}\,\mathrm{erg\,s^{-1}\,cm^{-2}\,\Angstrom^{-1}}$. Surface brightness in magnitudes per square arcsecond was then computed as
\begin{equation}
    \mu=-2.5\log_{10}\!\left(\frac{\mathrm{PHOTFLAM}\times{I}}{p^2}\right)+\mathrm{PHOTZPT},
\end{equation}
    \noindent where $I$ is the measured intensity and $p$ is the pixel scale in $\mathrm{arcsec\,pixel^{-1}}$.
    
    The limited field of view of the \textit{HST}/WFPC2 PC1 imaging complicates direct sky background estimation. To address this, we matched the outer \textit{HST} profile to the CGS \citep{Ho2011} $V$-band profile using a $\chi^2$ minimization over radii of $1$--$10\arcsec$, applying the resulting magnitude offset ($\simsym1.6$ mag) as a global sky correction. The final product is a sky-corrected, flux-calibrated $V$-band surface brightness profile for NGC~5102.

    Next, to characterize the nuclear star cluster (NSC), bulge, and disk components of NGC~5102, we fit the surface brightness profile with a four-component Sérsic model, consisting of an NSC, two bulge-like components, and a disk, using a bounded, weighted non-linear least-squares algorithm. The fit weights are inversely proportional to the observed intensity at each radius. Parameter uncertainties are estimated from the Jacobian-based covariance of each least-squares solution and propagated with Monte Carlo sampling to obtain radial uncertainty bands for each component and for the total model profile. Figure~\ref{fig:sb} presents a summary of the surface brightness profile analysis, showing the measured, azimuthally averaged surface-brightness profile from the \textit{HST} image and CGS data along with the decomposition and residuals. The best-fit Sérsic parameters and their $1\sigma$ uncertainties are listed in Table~\ref{tab:sersic_params}.

\begin{figure*}
    \centering
    \includegraphics[width=0.8\paperwidth]{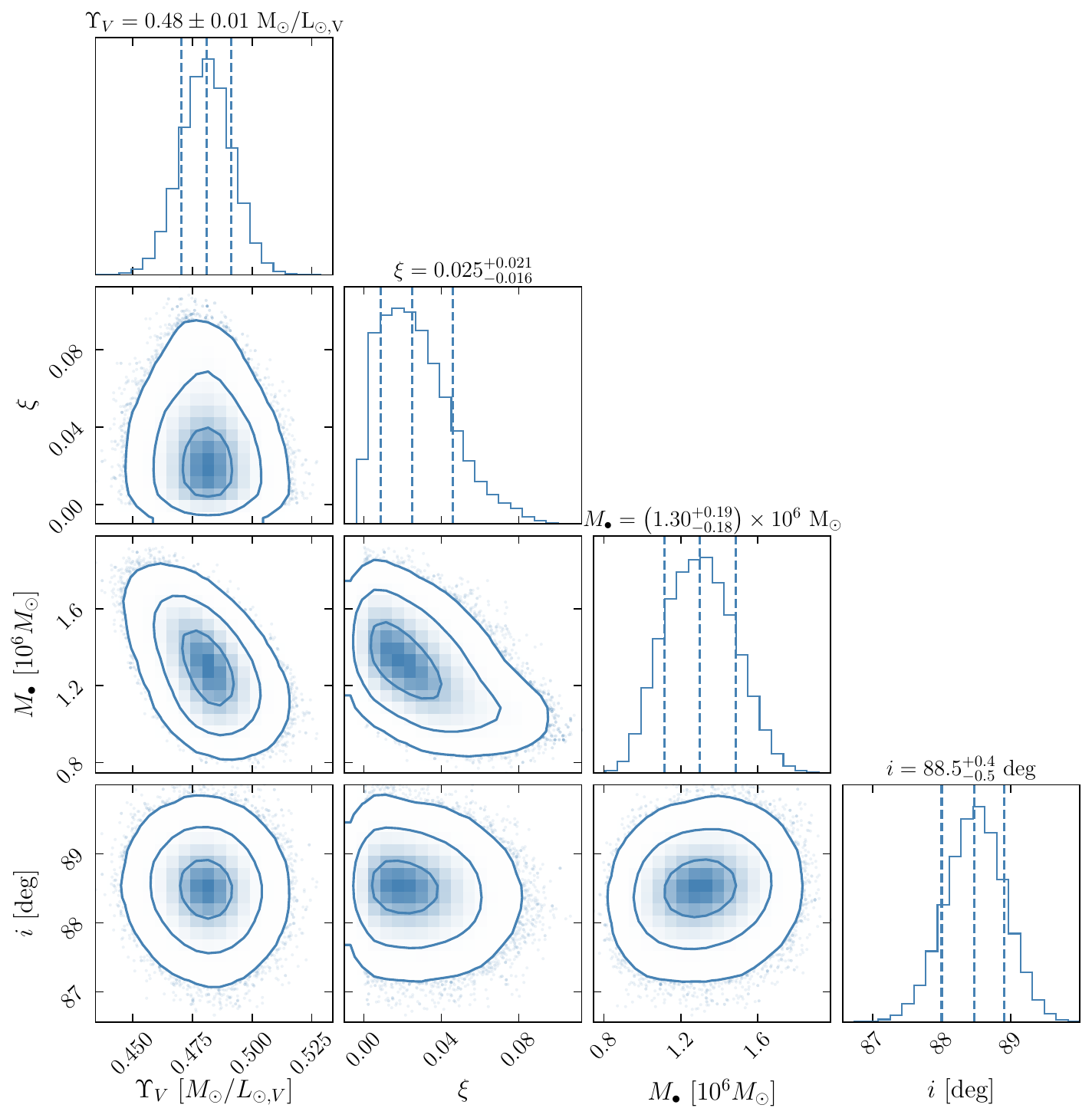}
    \caption{Corner plot showing the marginalized posterior distributions for the four free parameters in our SCO modeling framework: bulge dynamical $M/L$ $\Upsilon_V$, NSC-to-bulge $M/L$ $\xi$, black hole mass $\Mbh$, and inclination $i$. Contours enclose the 68.3\%, 95.4\%, and 99.7\% posterior credible regions. The black hole mass is well constrained at $\Mbh=(1.30^{+0.19}_{-0.18})\times10^6\,\Msun$. We quantify parameter covariances using Pearson correlation coefficients, $r_{XY}\equiv\mathrm{cov}(X,Y)/(\sigma_X\sigma_Y)$, computed from the posterior samples. The posterior distributions exhibit moderate anti-correlations between $\Mbh$ and $\Upsilon_V$ ($r_{M\Upsilon}=-0.52$) and between $\Mbh$ and $\xi$ ($r_{M\xi}=-0.56$). The inclination shows a weaker anti-correlation with $\xi$ ($r_{i\xi}=-0.48$). These covariances are visible as tilted joint posteriors in the corresponding panels. All marginalized posterior distributions are unimodal, indicating that each parameter is well constrained.}
    \label{fig:corner}
\end{figure*}

\section{Kinematic Extraction: Line-of-Sight Velocity Distribution Fitting} \label{sec:kinematic_extract}

    To extract stellar kinematics for NGC~5102, we used MUSE WFM IFU and \textit{HST}/STIS spectra. Prior to extracting the line-of-sight velocity distributions (LOSVDs), we determined the kinematic major axis position angle (PA) using the \texttt{PaFit} software package \citep{Krajnovic2006}, which fits bisymmetric velocity fields and minimizes $\chi^2$ as a function of position angle. We fit a parabola to the global minimum and estimate $1\sigma$ errors from the zero crossings where $\Delta\chi^2=1$, yielding a kinematic PA of $45.77 \pm 0.23$ degrees, measured east of north.

    Following this determination, the spectral data were binned into 20 logarithmically spaced radial bins and five angular bins distributed equally in $\sin(\theta)$, producing finer bins near the kinematic major axis. Alternative adaptive binning schemes, such as Voronoi tessellation \citep[e.g.,][]{Cappellari2003} and its power-diagram generalizations, including PowerBin/centroidal power diagrams \citep[e.g.,][]{Cappellari2025}, are commonly used for integral-field data because they can provide spatially adaptive sampling at approximately fixed signal-to-noise. For the present analysis, however, we adopted a radial--angular binning scheme because it is matched to the axisymmetric geometry assumed in the Schwarzschild modeling and provides direct control over the sampling of the nuclear region and kinematic major axis. This binning also keeps the number of spatial apertures computationally tractable for the SCO modeling, where each additional aperture adds LOSVD constraints that must be evaluated for every orbit library and trial potential. The adopted bins provide sufficient signal-to-noise for robust LOSVD extraction while preserving the radial and angular information most relevant for constraining the central potential.
    
    The same spatial framework was then used to incorporate the STIS spectra. Because the STIS slit was aligned with the photometric major axis, which is offset from the kinematic major axis, the STIS spectra were assigned to the appropriate angular bin ($11.5$--$23.5$ degrees). As the galaxy displays uniform concentric elliptical isophotes (see Figure~\ref{fig:images}), we assume an axisymmetric system. We then reflected the spectra across the kinematic major axis and combined them in their corresponding bins. The spectral modeling framework for each spatial bin proceeds in two stages, described below: an Asymmetric Least Squares (AsLS) continuum fit and normalization followed by penalized-likelihood LOSVD extraction via template convolution.

\subsection{Continuum Normalization}

    We performed continuum fitting around the \CaII\ triplet region ($\simsym 8400$--$8750\,\Angstrom$) using an AsLS algorithm \citep{Eilers1996, Eilers2003}. This robust baseline-fitting approach iteratively weights data points based on residuals from the fitted continuum, allowing the continuum to be estimated while reducing the influence of absorption and emission features. This asymmetric weighting naturally isolates the continuum while preserving stellar absorption features. Prior to continuum fitting, we performed automated detection and masking of:

    \begin{enumerate}
        \item Absorption regions: Identified via localized deficits exceeding $5\sigma$ below the smoothed spectrum and expanded by 5 pixels to capture extended absorption wings.
        \item Emission regions: Identified as localized excesses exceeding $5\sigma$ above the smoothed spectrum and expanded by 3 pixels.
        \item Residual sky features: Additional features identified from fitting residuals.
    \end{enumerate}

    All masked regions were excluded from the continuum fit. To prevent continuum bias, the AsLS algorithm was applied only to the unmasked (good) pixels. The resulting continuum estimate was then interpolated across the masked regions. The normalized spectra were subsequently calculated as $f_{\mathrm{norm}}=f_{\mathrm{obs}}/f_{\mathrm{continuum}}$, with error propagation handled according to the input noise estimates.

\subsection{Stellar Template Fitting and LOSVD Extraction}

    The kinematics were then extracted using a maximum penalized likelihood approach \citep{Gebhardt2000, Merritt1997, Saha1994}, as implemented in the SCO modeling framework. The continuum-normalized spectra in each bin were fitted as a non-negative linear combination of stellar templates drawn from the MUSE stellar library \citep{Ivanov2019} as well as representative stellar spectra observed with STIS. Specifically, G and K giant star templates were used, as they dominate the galaxy luminosity in the optical and infrared and display strong \CaII\ absorption features characteristic of evolved stellar populations.

    The templates were convolved with a non-parametric LOSVD represented on a discretized velocity grid with uniform spacing based on $N_v = 29$ bins spanning $\pm 4.5\,\sigma_0$, where $\sigma_0$ is the initial dispersion estimate for the spectrum being fitted. The corresponding bin spacing is $\Delta v = 9\,\sigma_0/(N_v-1)$, ranging from ${\approx}16$ to $26\,\kms$ across the spectra used here. This grid was set wide enough to capture stellar motions well into the wings of the distribution, while the bin spacing is finer than the instrumental resolution elements. We estimate these instrumental velocity resolutions from the spectral FWHM values described in Section~\ref{sec:observations} using $c\,\Delta\lambda/\lambda$, giving ${\sim}88\,\kms$ for MUSE ($\Delta\lambda \simsym 2.5\,\Angstrom$) and ${\sim}40$--$55\,\kms$ for STIS ($\Delta\lambda \simsym 1.1$--$1.6\,\Angstrom$) near $8500\,\Angstrom$. Thus, the LOSVD extraction grid provides adequate velocity sampling without introducing an excessive number of free parameters, and the resulting model spectrum was directly compared to the continuum-normalized observed spectrum. Absorption and emission lines beyond the \CaII\ triplet region, as well as any residual sky features and detector artifacts, were masked from the fit using a bad pixel mask. Spectra with excessive numbers of zero-valued pixels were automatically identified and excluded from further analysis.

    The parameter estimation minimized a weighted $\chi^2$-like objective function incorporating a regularization term that penalizes curvature (second derivatives) in the LOSVD. In practice, the extraction used an equal-weight data-misfit term over the unmasked continuum-normalized pixels, together with smoothness and normalization penalties. The objective function is
\begin{align}
    F &= \chi^2 + P_{\rm smooth} + P_{\rm norm} \nonumber \\
      &= \sum_{i} \left[g_{\rm obs}(x_i) - g_{\rm model}(x_i)\right]^2 \nonumber \\
      &\quad + \sum_{j=2}^{N_v-1} \lambda_j \bigl(b_{j+1} - 2b_j + b_{j-1}\bigr)^2 \nonumber \\
      &\quad + 0.1\,\bigl|\textstyle\sum_j b_j - 1\bigr| \,,
    \label{eq:fitlov_obj}
\end{align}
    \noindent where $g_{\rm obs}$ is the continuum-normalized observed spectrum, $g_{\rm model}$ is the template combination convolved with the trial LOSVD, and $b_j$ are the LOSVD bin values. Equivalently, this corresponds to setting $\sigma_i = 1$ for all unmasked pixels in the $\chi^2$-like term; this choice is appropriate because the continuum-normalized spectra have approximately uniform signal level across the narrow fitting window. The regularization weight $\lambda_j$ is position-dependent: $\lambda_j = \lambda$ in the velocity core ($|v_j| \leq 1.8\,\sigma_0$), rising steeply as $\lambda_j = \lambda (|v_j|/1.8\,\sigma_0)^4$ in the wings ($|v_j| > 1.8\,\sigma_0$). This design applies uniform regularization in the core, where the LOSVD signal is strong and the shape is well constrained by the data, while more strongly suppressing nonphysical structure in the low-signal wing region. The global regularization strength $\lambda$ was selected by comparing fits over a range of values and choosing the smallest $\lambda$ that produced smooth, physically plausible LOSVDs without noticeably degrading the spectral residuals; this criterion balances fidelity to the data against artificial over-smoothing. Normalization constraints were applied to ensure physical consistency for both the LOSVD and the template fractions, enforcing positivity and unit-sum properties as appropriate. Specifically, both the LOSVD bins ($b_j \geq 10^{-6}$) and the template mixing weights ($w_k \geq 10^{-5}$) were constrained to be non-negative. The third term in Equation~\ref{eq:fitlov_obj} is a soft penalty that encourages unit normalization of the LOSVD; the coefficient $0.1$ was chosen to be large enough to enforce near-normalization without dominating the data-misfit term. The minimization was performed using a bound-constrained optimizer.

    Uncertainties in the inferred kinematic parameters were quantified via Monte Carlo methods: synthetic spectra were generated by adding per-pixel Gaussian noise, with amplitudes set by the smoothed fit residuals of the best-fit solution, to the best-fit model spectrum. Each synthetic spectrum was independently refitted using the same parameter-estimation procedure. This empirical residual-based Monte Carlo procedure generated an ensemble of LOSVDs, template weights, and Gauss-Hermite moment measurements. Central values for all quantities are the penalized-likelihood best-fit values from the unperturbed spectrum; confidence limits are derived from the Monte Carlo ensemble using a biweight location estimator to robustly anchor the percentile offsets (16th and 84th percentiles, corresponding to $\simsym 1\sigma$ errors for Gaussian distributions). To enable a compact kinematic summary and facilitate comparison with the dynamical models, each LOSVD realization was additionally characterized by fitting a Gauss-Hermite series \citep{vanderMarel1993, Gerhard1993}, yielding estimates of the mean velocity $V$, velocity dispersion $\sigma$, and the higher-order moments $h_3$ and $h_4$, which describe asymmetric and symmetric deviations from a Gaussian, respectively. The Schwarzschild modeling uses the full non-parametric LOSVDs rather than only these Gauss-Hermite summaries. Regions consistently affected by sky residuals or detector artifacts were identified and excluded from the fitting.

    A summary of the spectral fitting is shown in Figure~\ref{fig:spectra}, which presents representative central spectra for MUSE and STIS before and after continuum normalization together with the best-fit template combinations. For comparison with the Schwarzschild orbit library, the extracted non-parametric LOSVDs were interpolated onto the model velocity grid ($\Delta v = 50\,\kms$, spanning $\pm300\,\kms$). Examples of these LOSVDs along the kinematic major axis, compared to the best-fit dynamical model, are shown in Figure~\ref{fig:losvd}. Finally, Figure~\ref{fig:kinematics} summarizes the extracted one-dimensional kinematic profiles, comparing the mean velocity $v$ and velocity dispersion $\sigma$ as a function of radius for STIS and MUSE. The higher spatial resolution of \textit{HST}/STIS resolves a stronger central rise in $\sigma$ than is seen with MUSE, reflecting superior resolution near the sphere of influence; the MUSE profiles provide complementary constraints at larger radii. The lower central $\sigma$ measured with MUSE is expected given its coarser spatial sampling relative to the estimated sphere of influence, so the central dispersion peak is partially suppressed by beam smearing.

\section{Modeling} \label{sec:modeling}

    Here, we outline our SCO modeling framework. We first discuss our model for a radially varying $M/L$ profile to account for the presence of a central, blue nuclear star cluster. Next, we describe our SCO modeling methodology in detail. Finally, we present the results of our modeling of NGC~5102.

\subsection{Variable Mass-to-Light Ratio} \label{sec:vml}

    NGC~5102 displays prominent radial color gradients and evidence of a distinct stellar population in its nucleus, characterized by a blue, compact NSC embedded within an older, redder bulge \citep{Nguyen2018, Nguyen2019}. The presence of young, massive stars in the NSC lowers the stellar $M/L$ ($\Upsilon$) at small radii relative to the bulge. Accurate modeling of these spatial variations in $\Upsilon$ is therefore essential for any dynamical analysis seeking to recover the central black hole mass, because assuming a constant $\Upsilon$ would systematically overestimate the stellar mass in the nucleus and therefore underestimate the mass of the central MBH.
    
    To address this, we construct a radially varying $M/L$ profile using our decomposition of the observed one-dimensional surface brightness profile (Section~\ref{sec:surface_brightness}). Because our decomposition is performed in projection, we construct an effective, radially varying $M/L$ profile $\Upsilon(r)$ as the ratio of the stellar surface mass density $\Sigma_M(r)$ (in units of $\Msun\,\mathrm{pc}^{-2}$) to the stellar surface brightness $\Sigma(r)$ (in units of $\Lsun\,\mathrm{pc}^{-2}$):
\begin{equation} \label{eq:derv_step1}
    \Upsilon(r) \equiv \frac{\Sigma_M(r)}{\Sigma(r)}.
\end{equation}
    
    We then express the total surface brightness and surface mass density as the sum of the bulge and NSC components from the decomposition, $\Sigma(r)=\Sigma_{\mathrm{B}}(r)+\Sigma_{\mathrm{NSC}}(r)$ and $\Sigma_M(r)=\Sigma_{M,{\mathrm{B}}}(r)+\Sigma_{M,{\mathrm{NSC}}}(r)$, yielding
\begin{equation} \label{eq:derv_step2}
    \Upsilon(r)=\frac{\Sigma_{M,{\mathrm{B}}}(r)+\Sigma_{M,{\mathrm{NSC}}}(r)}{\Sigma_{\mathrm{B}}(r)+\Sigma_{\mathrm{NSC}}(r)}, 
\end{equation}
    \noindent where $\Sigma_{\mathrm{B}}(r)$ and $\Sigma_{\mathrm{NSC}}(r)$ are the bulge and NSC surface brightness profiles, and $\Sigma_{M,{\mathrm{B}}}(r)$ and $\Sigma_{M,{\mathrm{NSC}}}(r)$ are the corresponding stellar surface mass densities.

    Next, we assume that each component has a constant intrinsic $M/L$, denoted by $\Upsilon_{\mathrm{B}}$ and $\Upsilon_{\mathrm{NSC}}$ for the bulge and NSC, respectively, such that
    $\Sigma_{M,{\mathrm{B}}}(r)=\Upsilon_{\mathrm{B}}\Sigma_{\mathrm{B}}(r)$ and
    $\Sigma_{M,{\mathrm{NSC}}}(r)=\Upsilon_{\mathrm{NSC}}\Sigma_{\mathrm{NSC}}(r)$.
    Substituting into Equation~(\ref{eq:derv_step2}) yields
\begin{equation} \label{eq:derv_step3}
    \Upsilon(r) =
    \frac{\Upsilon_{\mathrm{B}}\Sigma_{\mathrm{B}}(r)+\Upsilon_{\mathrm{NSC}}\Sigma_{\mathrm{NSC}}(r)}
         {\Sigma_{\mathrm{B}}(r)+\Sigma_{\mathrm{NSC}}(r)}.
\end{equation}

    Finally, we parameterize the relative normalization by introducing $\xi \equiv \Upsilon_{\mathrm{NSC}}/\Upsilon_{\mathrm{B}}$, so that
\begin{equation} \label{eq:vml}
    \Upsilon(r)=\Upsilon_{\mathrm{B}}
    \left[
    \frac{\Sigma_{\mathrm{B}}(r)+\xi\,\Sigma_{\mathrm{NSC}}(r)}
         {\Sigma_{\mathrm{B}}(r)+\Sigma_{\mathrm{NSC}}(r)}
    \right].
\end{equation}
    
    The final radially varying $M/L$ model is parameterized by $\Upsilon_{\mathrm{B}}$ (hereafter denoted as $\Upsilon_V$ because our surface-brightness profile is in the $V$ band) and $\xi$, both of which are allowed to vary in our Schwarzschild orbit modeling (Section~\ref{sec:sco_modeling}). Figure~\ref{fig:vml} illustrates the radially varying $\Upsilon(r)$ profile recovered from our best-fit dynamical model and demonstrates a pronounced drop in $\Upsilon(r)$ toward the galaxy center, consistent with the blue colors and recent star formation activity in the NSC. Explicitly modeling this spatial variation in $\Upsilon$ mitigates systematic errors associated with assuming radially uniform $M/L$, particularly in galaxies where nuclear star clusters are present. Incorporating the variable $M/L$ into our dynamical framework improves the stellar mass estimates and, crucially, reduces systematic bias in the measurement of the central MBH mass in NGC~5102.

\subsection{Schwarzschild Orbit-superposition Modeling Framework} \label{sec:sco_modeling}

    To infer the mass of the central black hole in NGC~5102, we employ axisymmetric, three-integral Schwarzschild orbit-superposition models \citep{Schwarzschild1979}. Our implementation closely follows the methodology established by \citet{Gebhardt2000, Gebhardt2003, Siopis2009}; this approach has also been used in more recent works \citep[e.g.,][]{Gultekin2024, Waters2024, Lujan2025}. We modify this framework to incorporate a radially varying $M/L$ (Section~\ref{sec:vml}), joint STIS+MUSE constraints, and an expanded posterior inference framework.

    In the implemented framework, each point in parameter space is evaluated through a consistent sequence of steps: constructing the deprojected mass model, generating an orbit library in the trial potential, projecting model observables to the line of sight, convolving with the instrument response, and solving for non-negative orbital weights that best reproduce the photometric and kinematic constraints. This end-to-end sequence is repeated for each point in parameter space and yields the model-comparison statistics used for inference.

    First, we determine the stellar surface-brightness profile using the XVISTA software and \textit{HST}/WFPC2 imaging data (Section~\ref{sec:surface_brightness}). This profile is deprojected under the assumption of axisymmetry and an inclination $i$, yielding an intrinsic luminosity density $\nu(r,\theta)$, where $\theta$ is the polar angle. We then construct a radially varying stellar mass-to-light ratio $\Upsilon(r)$ using the bulge+NSC decomposition (see Section~\ref{sec:vml}) and convert the luminosity density to a stellar mass density via $\rho_\ast(r,\theta)=\Upsilon(r)\,\nu(r,\theta)$ (with $\Upsilon$ independent of $\theta$). The gravitational potential is then computed by solving Poisson's equation for the stellar mass distribution and is augmented with a central point mass representing the MBH, parameterized by its mass ($\Mbh$).

    Given a set of trial parameters ($\Mbh$, $\Upsilon_V$, $\xi$, $i$), we generate a library of $\simsym10^3$--$10^4$ stellar orbits sampling the available phase space. For each orbit, we track the time spent in each projected spatial bin; the model surface brightness in each bin is then computed as the weighted sum of these time-in-bin contributions across all orbits, with weights determined by the non-negative least-squares fit. The orbital weights are optimized such that the model reproduces both the observed surface brightness and the LOSVDs extracted from the STIS and MUSE spectra (Section~\ref{sec:kinematic_extract}). Prior to data-model comparison, model predictions are convolved with the instrumental PSFs. The PSFs used in the dynamical modeling were implemented as circular two-dimensional Gaussian kernels. To estimate the adopted widths, we identified isolated, unsaturated point sources in the MUSE and \textit{HST} fields after estimating the local background with sigma-clipped statistics. Centered cutouts of these point sources were median-combined to construct empirical PSFs, and an effective FWHM was measured from the azimuthally averaged radial profiles. We then adopted Gaussian PSFs with $\mathrm{FWHM}=0\farcs42$ for MUSE and $\mathrm{FWHM}=0\farcs09$ for \textit{HST}/STIS in the Schwarzschild modeling.
    
    The parameter space was first explored with a coarse grid over plausible values of ($\Mbh$, $\Upsilon_V$, $\xi$, $i$), followed by finer grid refinement around the global $\chi^2$ minimum. The four free parameters allow the model to account simultaneously for inclination and radially varying stellar $M/L$ induced by the NSC. Fit quality is quantified with $\chi^2$ statistics comparing model and observed LOSVD and surface brightness constraints.
    
    For posterior inference, we use the resulting model grid as input to a likelihood-sampling stage. Specifically, we compile the full grid of model outcomes, apply a high-$\chi^2$ cutoff for numerical stability, and interpolate the retained 4D $\chi^2$ surface with a scaled radial-basis-function (RBF) interpolator. In the fiducial setup, we use a multiquadric RBF kernel with finite smoothing and evaluate
\begin{equation}
    \ln \mathcal{L}(\boldsymbol{\theta})=-\frac{1}{2}\chi^2_{\mathrm{interp}}(\boldsymbol{\theta}),
\end{equation}
    within the sampled parameter bounds.
\begin{figure*}
    \centering
    \includegraphics[width=0.8\paperwidth]{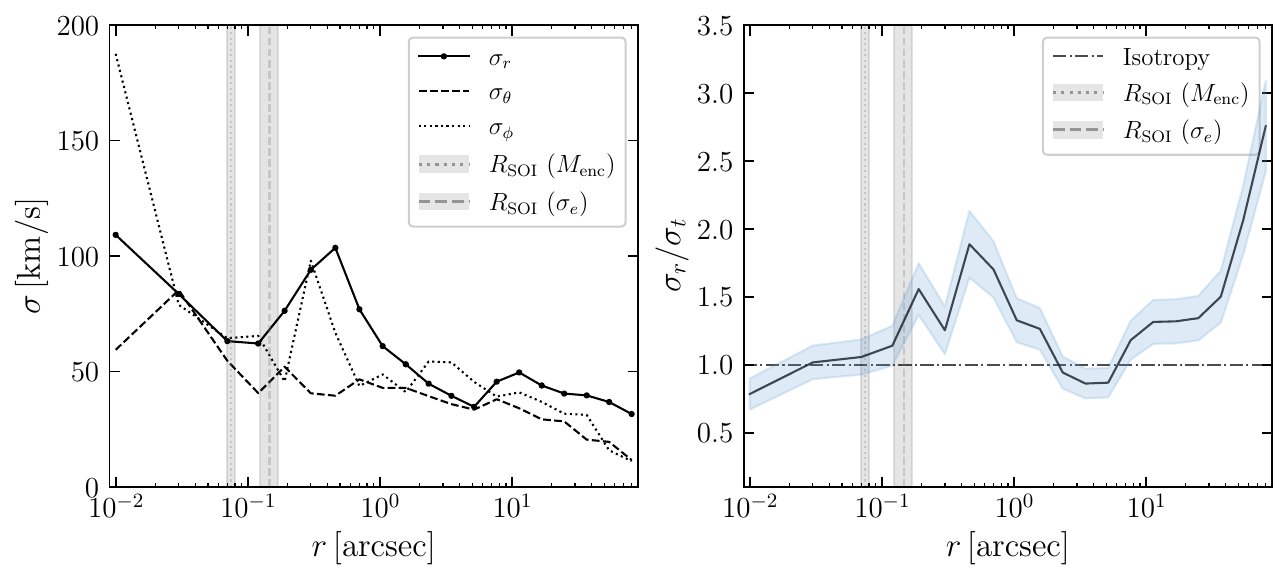}
    \caption{Velocity dispersion radial profiles from our best-fit SCO model decomposed into the radial component $\sigma_r$, the polar component $\sigma_\theta$, and the azimuthal component $\sigma_\phi$. Gray shaded regions mark $R_{\mathrm{SOI}}$ inferred from $M_{\mathrm{enc}}$ (dotted) and $\sigma_e$ (dashed). The right panel shows the orbital anisotropy, the ratio of the radial velocity dispersion ($\sigma_r$) to the tangential velocity dispersion ($\sigma_t$) as a function of radius, where $\sigma_t^2 \equiv 0.5(\sigma_\theta^2+\sigma_\phi^2)$, so that $\sigma_r/\sigma_t=1$ for isotropy. The blue shaded region shows $1\sigma$ uncertainties. The anisotropy profile suggests a complex structure: the inner $\simsym0\farcs03$ (the NSC) is tangentially biased, consistent with a rotating or disk-like component. Between $\simsym0\farcs03$ and $\simsym2\arcsec$ the orbits become radially biased, indicating a transition to a more dispersion-supported spheroidal component. The profile then dips to tangential bias between $\simsym2\arcsec$ and $\simsym6\arcsec$, possibly signaling another rotationally supported or isotropic component, and rises sharply at larger radii to strong radial bias. The initial transition between tangential and radial bias occurs near $R_{\mathrm{SOI}}$ (from $M_{\mathrm{enc}}$) and the effective radius $R_e$ of the NSC, suggesting that the gravitational potential of the central SMBH dominates the kinematics of the NSC.}
    \label{fig:sig_profiles}
\end{figure*}   
 
    We then sample this likelihood surface with an ensemble MCMC sampler. The default analysis adopts flat priors within the allowed bounds (with optional informative priors explored as a consistency check), and uses a mixed move set (StretchMove + Differential Evolution moves), 32 walkers, and long chains with explicit burn-in and thinning. Because interpolated likelihood surfaces can contain sparse low-probability pockets, we apply robust post-processing to reject poorly converged walkers and extremely low-probability samples before computing final credible intervals. This framework yields robust constraints on $\Mbh$, the stellar mass normalization, and orientation while explicitly propagating the impact of the variable $M/L$ profile.
 
\subsection{Modeling Results} \label{sec:modeling_results}

    We constrained the central black hole mass in NGC~5102 with axisymmetric SCO models that simultaneously fit the \textit{HST}/STIS and VLT/MUSE LOSVDs (Section~\ref{sec:kinematic_extract}) and the \textit{HST}/WFPC2-based surface brightness model (Section~\ref{sec:surface_brightness}). The dynamical model includes a radially varying stellar $M/L$ (Section~\ref{sec:vml}) parameterized by a bulge $V$-band $M/L$ ratio, $\Upsilon_V$, and an NSC-to-bulge $M/L$, $\xi \equiv \Upsilon_{\mathrm{NSC}}/\Upsilon_V$, and allows the galaxy inclination $i$ to vary.
    
    We explored the four-dimensional parameter space $\boldsymbol{\theta}\equiv(\Upsilon_V,\,\xi,\,\Mbh,\,i)$ over the bounds
    \begin{equation}
    \begin{aligned}
        \Upsilon_V &\in [0.1,\,2.0]\,\MsunLsun,\\
        \xi &\in [0.00,\,1.0],\\
        \Mbh &\in [0,\,3.0\times10^{6}]\,\Msun,\\
        i &\in [70^\circ,\,90^\circ].
    \end{aligned}
    \end{equation}
    These bounds were chosen to be broad enough to contain physically plausible solutions while avoiding regions of parameter space that are either unphysical or computationally uninformative. They were also guided in part by the previous dynamical modeling of NGC~5102 by \citet{Nguyen2018, Nguyen2019}, but were intentionally chosen to be substantially broader than the reported uncertainties from that work. The lower bound $\Mbh=0$ explicitly permits a no-black-hole model, while the upper bound $3.0\times10^6\,\Msun$ is several times larger than both the previous JAM-based estimate for NGC~5102 and the value expected from standard $\Mbh$--$\sigma_e$ extrapolations, ensuring that the allowed range brackets any plausible central mass. The stellar mass-to-light ratio range $\Upsilon_V\in[0.1,2.0]\,\MsunLsun$ spans a conservative range around the stellar $M/L$ values inferred by \citet{Nguyen2018, Nguyen2019} and allows for uncertainties associated with the young/intermediate-age stellar populations in NGC~5102, while enforcing positive stellar mass. The NSC-to-bulge ratio $\xi\in[0,1]$ encodes the assumption, motivated by the blue nuclear stellar population, that the NSC has a stellar $M/L$ less than or equal to that of the surrounding bulge; the lower limit allows the limiting case of a negligibly small NSC stellar mass contribution. Finally, the inclination range $i\in[70^\circ,90^\circ]$ reflects the highly inclined morphology of NGC~5102 and includes the edge-on limit, while excluding substantially more face-on deprojections that are inconsistent with the observed flattening and with previous inclination estimates. We adopt flat priors within these bounds.

    We evaluated approximately $3.2\times10^{4}$ SCO models on a grid. The resulting $\chi^2$ grid was processed via the likelihood-sampling pipeline described in Section~\ref{sec:sco_modeling}. In the implemented workflow, we assess sampling quality with acceptance-fraction and autocorrelation diagnostics and then derive posterior constraints from post-burn-in, thinned chains. We additionally evaluate robustness to numerical choices by varying the high-$\chi^2$ retention cut and low-log-probability filtering thresholds used in chain cleaning. These checks preserve the qualitative posterior structure and keep the inferred $\Mbh$ within the quoted credible intervals. Except for the physically motivated non-negativity limit on $\xi$, the marginalized posterior distributions are contained well within the adopted bounds, indicating that the inferred parameters are constrained primarily by the data and model likelihood rather than by the imposed parameter limits.

    Central values are taken as posterior medians, with credible intervals from percentile-based summaries (68.3\%, 95.4\%, and 99.7\% equivalents). The marginalized posteriors (Figure~\ref{fig:corner}) yield
    \begin{align*}
        \Mbh &=(1.30^{+0.19}_{-0.18}\,(^{+0.36}_{-0.32},\,^{+0.50}_{-0.43}))\times10^6\,\Msun,\\
        \Upsilon_V  &=0.48 \pm 0.01\,(0.02,\,0.03) \,\MsunLsun,\\
        \xi         &=0.025^{+0.021}_{-0.016}\,(^{+0.048}_{-0.024},\,^{+0.068}_{-0.022}),\\
        i           &=88.5^{+0.4}_{-0.5}\,(^{+0.8}_{-0.9},\,^{+1.2}_{-1.4})\,\mathrm{deg},
    \end{align*}
    where quoted uncertainties are 68.3\% (95.4\% and 99.7\%) credible intervals. The corresponding one- and two-dimensional marginalized posteriors are shown in Figure~\ref{fig:corner}.
    
    A model with no central black hole ($\Mbh=0$) is also strongly disfavored. Comparing the best-fit $\Mbh=0$ model to the best-fit free-$\Mbh$ model yields $\Delta\chi^2=90$, indicating that reallocating mass among stellar components alone cannot reproduce the observed nuclear LOSVD widths without an additional compact central mass. This comparison was performed with a dedicated fixed-$\Mbh$ likelihood analysis (sampling $(\Upsilon_V,\xi,i)$ with $\Mbh=0$) and direct comparison to the free-$\Mbh$ solution. In the best-fit $\Mbh=0$ case, the model compensates by increasing the stellar nuclear mass, with the largest parameter shift in $\xi$, as expected from the degeneracy between nuclear stellar mass and black-hole mass.

    The inferred parameter covariances follow the expected central mass-partition degeneracies. We quantify these covariances using Pearson correlation coefficients, $r_{XY}\equiv\mathrm{cov}(X,Y)/(\sigma_X\sigma_Y)$, computed from the posterior samples. In particular, $\Mbh$ is anti-correlated with $\Upsilon_V$ ($r_{M\Upsilon}=-0.52$) and with $\xi$ ($r_{M\xi}=-0.56$): increasing stellar mass normalization, either globally via $\Upsilon_V$ or centrally via $\xi$, reduces the required black-hole mass to reproduce the inner LOSVD widths. The inclination is comparatively weakly coupled to $\Mbh$ over the explored range, although it shows a modest anti-correlation with $\xi$ ($r_{i\xi}=-0.48$).

    The credible intervals quoted above describe the statistical uncertainty within our adopted SCO model. They do not include all possible systematic uncertainties. In particular, stellar-dynamical black-hole masses can depend, for example, on assumptions about the galaxy geometry, the instrumental PSF, stellar templates, and stellar $M/L$ model; such effects are known to be important in orbit-superposition measurements and can be comparable to, or larger than, the formal statistical errors \citep{Gebhardt2009, Nguyen2018, Nguyen2019, Schulze2011, Valluri2004, vandenBosch2010}. Therefore, while the large $\Delta \chi^2$ relative to $\Mbh = 0$ supports the detection of a compact central mass in our adopted modeling framework, comparisons to other methods or to scaling relations should account for this broader systematic uncertainty.

\begin{figure*}
    \centering
    \includegraphics[width=0.8\paperwidth]{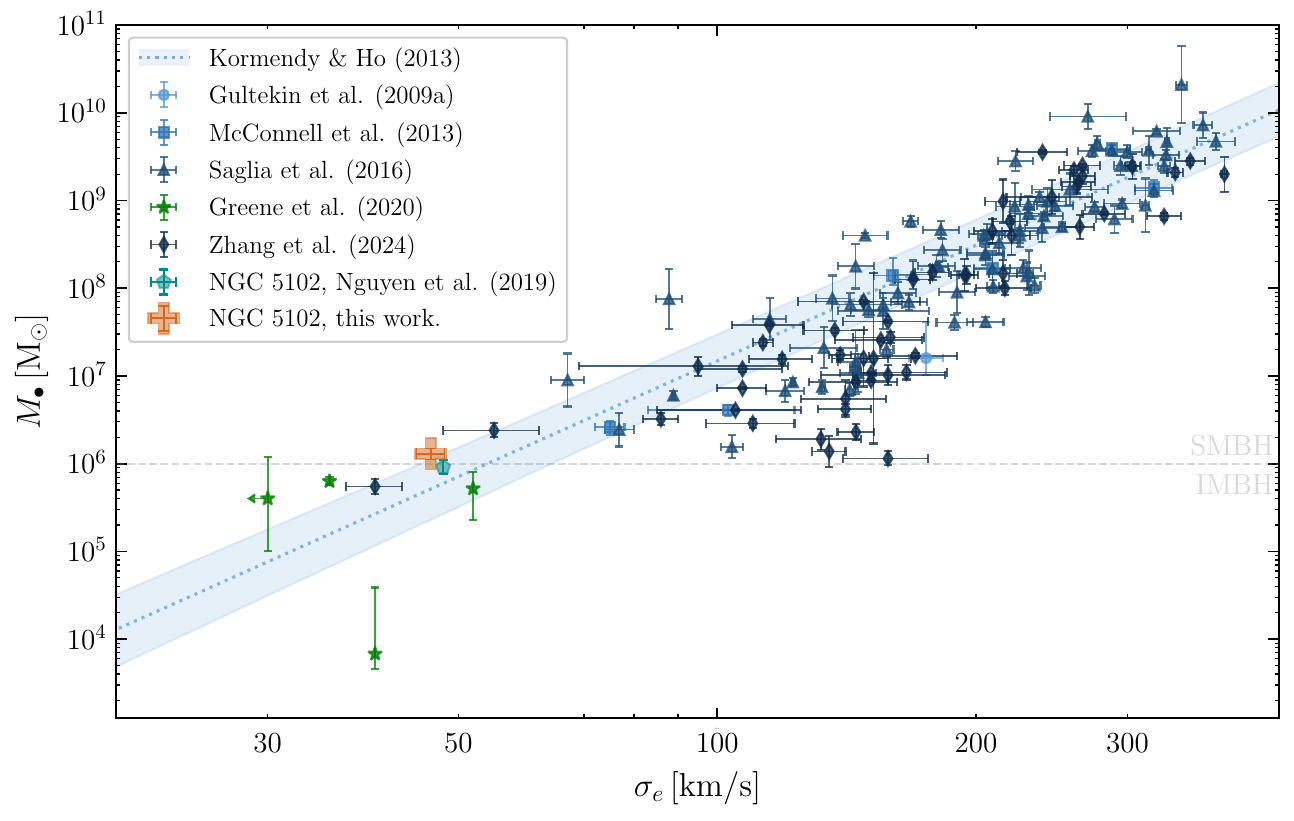}
    \caption{Position of NGC~5102 on the Kormendy \& Ho (2013) $\Mbh$--$\sigma_e$ relation (blue dotted line). The uncertainties of the $\Mbh$--$\sigma_e$ relation, plotted as the blue filled band, include both the individual parameter uncertainties (where $\alpha=0.309^{+0.037}_{-0.033}$, $\beta=4.38\pm0.29$; see Equation~\ref{eq:KH2013}) and the intrinsic scatter ($0.29$ dex). Currently available dynamically measured MBH masses are shown as data points with varying marker types and a blue color scale: Gültekin et al. (2009a, light blue circles), McConnell \& Ma (2013, medium blue squares), Saglia et al. (2016, dark blue triangles), and Zhang et al. (2024, navy blue diamonds). Also shown is the Greene et al. (2020) collection of dynamically measured IMBH candidates in galactic centers, shown as green stars. The cyan pentagon marks the previous NGC~5102 measurement by Nguyen et al. (2019), which found an IMBH of $(9.12^{+1.84}_{-1.53})\times10^5\,\Msun$. Our measurement (orange cross, $(1.30^{+0.19}_{-0.18})\times10^6\,\Msun$, within the SMBH regime) is $0.16$ dex larger than the Nguyen et al. (2019) measurement, but is consistent at the $\simsym1.7\sigma$ level. Relative to the $\Mbh$--$\sigma_e$ relation, our measured $\Mbh$ is higher than the predicted value of $(5.15^{+6.25}_{-2.92})\times10^5\,\Msun$ at $\sigma_e=46.4\pm1.7\,\kms$ by $0.40$ dex, but is consistent within the overall scatter of the relation.}
    \label{fig:msigma}
\end{figure*}   

    Finally, in addition to constraining global parameters, the orbit-superposition solutions provide diagnostics of internal structure. Figure~\ref{fig:sig_profiles} shows intrinsic dispersion components from the best-fit model ($\sigma_r$, $\sigma_\theta$, and $\sigma_\phi$) and the anisotropy profile $\sigma_r/\sigma_t$, with $\sigma_t^2 \equiv 0.5[\sigma_\theta^2+\sigma_\phi^2]$. The inner region is consistent with more tangentially biased orbital structure, transitioning outward toward more radial bias. We interpret this only qualitatively: the exact radii at which the profile changes are not uniquely determined, because nearby combinations of orbital weights and model assumptions can produce very similar anisotropy profiles. The robust conclusion is that the nucleus and surrounding bulge are dynamically distinct, with a multi-component orbital structure that cannot be described by a single, smoothly varying population.

\section{Discussion} \label{sec:discussion}

    Our primary result is a stellar-dynamical measurement of the central black hole mass in NGC~5102,
    $\Mbh=(1.30^{+0.19}_{-0.18})\times10^6\,\Msun$. In this section, we discuss (i) the kinematic evidence for a compact central mass, (ii) the role of the nuclear star cluster and the associated mass-partition degeneracy, and (iii) the comparison of our result with prior dynamical measurements and scaling-relation expectations.

\subsection{Kinematic evidence for a central black hole}

    The preference for a nonzero $\Mbh$ is driven by the nuclear kinematics at radii comparable to the black hole sphere of influence. The best-fit model reproduces the central LOSVDs measured with \textit{HST}/STIS (Figure~\ref{fig:losvd}) and the broader two-dimensional constraints from MUSE. The likelihood-ratio test against $\Mbh=0$ yields $\Delta\chi^2=90$, indicating that reallocating mass among stellar components alone cannot match the observed nuclear LOSVD widths.
    
    The comparison between STIS and MUSE kinematics (Figure~\ref{fig:kinematics}) also behaves as expected for a compact central mass: the STIS dispersion profile peaks more strongly in the central bin than the MUSE profile. Given that the MUSE spatial sampling is coarser than the enclosed-mass estimate of $R_{\mathrm{SOI}}$, beam smearing naturally suppresses the observed central dispersion in the MUSE data, while STIS resolves the rise more directly. This is precisely the regime in which combining high-resolution STIS constraints with wider-field MUSE constraints improves robustness: the former anchors the nuclear potential while the latter constrains the surrounding mass model and orbital structure.
    
    While the quoted credible interval on $\Upsilon_V$ is small, the practical uncertainty on $\Upsilon_V$ is likely dominated by systematics (e.g., imperfect axisymmetry, PSF characterization, template mismatch, or residual dust effects). Importantly, however, the black hole detection itself is not contingent on fine-tuning $\Upsilon_V$: the need for a compact central mass persists after marginalizing over $\Upsilon_V$, $\xi$, and $i$ (Figure~\ref{fig:corner}) and is reflected in the large $\Delta\chi^2$ relative to $\Mbh=0$.

\subsection{Role of the nuclear star cluster and variable $M/L$}

    One of the central motivations for this work is that many low-mass galaxies host distinct nuclear components (NSCs, young stellar populations, or bars), which can bias dynamical black hole measurements if they are not treated with appropriate stellar mass models (Section~\ref{sec:intro}). NGC~5102 is a clear example: its compact blue NSC means that the central few tens of parsecs are dynamically distinct from the surrounding bulge, so the inner potential should not be treated as a smooth extension of the outer galaxy.
    
    The NSC also affects how one should interpret the black hole sphere of influence. In a galaxy with a steep central rise in enclosed stellar mass, the enclosed-mass definition ($M_\ast(<r)=\Mbh$) can yield a smaller radius than the commonly used $\sigma_e$-based estimate $r_{\mathrm{SOI}}=G\Mbh/\sigma_e^2$. In NGC~5102, the rapid increase of $M_\ast(<r)$ through the NSC naturally pushes $r_{\mathrm{SOI},M_{\mathrm{enc}}}$ inward relative to $r_{\mathrm{SOI},\sigma}$ (Figures~\ref{fig:kinematics} and \ref{fig:vml}). This does not indicate inconsistency; it reflects that the nucleus is not well described as a smooth continuation of the bulge.

    More generally, the dynamical inference still contains a mass-partition degeneracy: in the central resolution elements, stellar mass normalization and black-hole mass can partially trade off against one another. The variable $M/L$ prescription reduces that freedom by tying the nuclear stellar mass to the observed NSC light profile, but it does not remove it entirely. This is why the black-hole constraint should be understood as a joint inference from the nuclear kinematics and the stellar mass model rather than as a direct measurement from the kinematics alone.
    
    Finally, the orbit-structure diagnostics (Figure~\ref{fig:sig_profiles}) suggest the nuclear dynamics are multi-component, with a tangentially biased inner region and more radially biased orbits outside the NSC. We interpret this only qualitatively: the exact radii at which the profile changes are not uniquely determined, because nearby combinations of orbital weights and model assumptions can produce similar anisotropy profiles. The robust conclusion is that low-mass galaxies with NSCs demand modeling frameworks that can accommodate realistic nuclear structure when extracting $\Mbh$.

\subsection{Comparison to prior work and scaling relations}

    NGC~5102 has a prior dynamical black hole measurement from \citet{Nguyen2018, Nguyen2019}, who used CO band-head kinematics with JAM and found $\Mbh=(9.1^{+1.8}_{-1.5})\times10^5\,\Msun$. Our Schwarzschild-based measurement is higher by 0.16 dex but consistent within $\simsym1.7\sigma$. Consistency between independent stellar-kinematic tracers---the \CaII\ triplet and CO band heads---and independent dynamical frameworks (Schwarzschild versus JAM) supports the robustness of the inferred mass scale and provides an external consistency check on JAM in this low-mass regime. In the context of the broader goals of this work (Section~\ref{sec:intro}), such cross-method agreement helps bracket systematic uncertainties that can become increasingly important at low masses.

    We compute the effective velocity dispersion, $\sigma_e$, within the projected half-light radius $R_e$ as
\begin{equation} \label{eq:sig_e}
    \sigma_e^2=\frac{\int_0^{R_e}\left[\sigma^2(r)+v^2(r)\right]\,I(r)\,dr}{\int_0^{R_e} I(r)\,dr},
\end{equation}
    where $\sigma(r)$ and $v(r)$ are the line-of-sight velocity dispersion and mean velocity profiles measured along the kinematic major axis, respectively, and $I(r)$ is the stellar surface brightness profile. Using the kinematic profiles from Section~\ref{sec:kinematic_extract}, we find $\sigma_e=46.4 \pm 1.7\,\kms$, where the quoted uncertainty is the $1\sigma$ confidence interval from Monte Carlo propagation.

    For reference, we adopt the KH13 $\Mbh$--$\sigma_e$ relation,
\begin{equation} \label{eq:KH2013}
    \frac{\Mbh}{10^9\,\Msun} =
    \left(0.309^{+0.037}_{-0.033}\right)
    \left(\frac{\sigma_e}{200\,\kms}\right)^{4.38 \pm 0.29},
\end{equation}
    which we use to compute the expected $\Mbh$ at the measured $\sigma_e$ for comparison in Figure~\ref{fig:msigma}. In the $\Mbh$--$\sigma_e$ plane, NGC~5102 lies $0.40$ dex above the KH13 ridgeline at $\sigma_e=46.4\pm1.7\,\kms$, but remains consistent once the intrinsic scatter is taken into account.

    More broadly, Figure~\ref{fig:msigma} suggests that the agreement with KH13 is tight at the high-dispersion end ($\sigma_e\gtrsim 190\,\kms$), while an apparent ``under-massive'' locus emerges at intermediate dispersions ($\sigma_e \sim 100$--$210\,\kms$), where a non-negligible subset of published dynamical masses fall below the KH13 ridgeline. At still lower dispersions ($\sigma_e\lesssim 70$--$80\,\kms$), the sample becomes sparse and the observed spread increases, limiting strong conclusions about whether the underlying relation changes form or whether the effect reflects population mix and selection.

    The intermediate-$\sigma_e$ under-massive locus is not uniformly distributed across host properties. Across multiple compilations, it is populated primarily by late-type galaxies, and barred spirals constitute a substantial fraction of the under-massive spiral subset. This pattern is qualitatively consistent with the expectation that non-axisymmetric structure can affect both the interpretation of $\sigma_e$ and the dynamical assumptions underlying some modeling approaches, potentially contributing to systematic offsets for particular host classes.

    Measurement selection may also influence the intermediate-$\sigma_e$ locus. In the \citet{Zhang2024} compilation, many of the under-massive spiral galaxies in this dispersion range have megamaser-based masses. Although maser measurements provide very small formal uncertainties, maser hosts are selected to contain thin, nearly edge-on circumnuclear disks with conditions favorable for masing \citep[e.g.,][]{Lo2005, Zhang2010}. This selection arises because detectable disk megamasers require long velocity-coherent path lengths through dense molecular gas. These requirements may correlate with specific host characteristics, nuclear gas content, AGN properties, or viewing geometries \citep[e.g.,][]{Greene2016}. The prominence of maser-selected spirals among the under-massive objects therefore motivates caution in interpreting the intermediate-$\sigma_e$ offset as a universal change in the $\Mbh$--$\sigma_e$ relation.

    As a simple descriptive check within our comparison set, we label a system ``under-massive'' if it lies below the KH13 relation by more than 0.4 dex. Under this definition, under-massive objects comprise $\simsym26\%$ of the full sample, with higher fractions among spirals ($\simsym48\%$), barred spirals ($\simsym50\%$), and megamaser measurements ($\simsym58\%$). These fractions are not a controlled statistical test---they depend on sample definition and on heterogeneous measurement uncertainties---but they support the qualitative conclusion that the intermediate-$\sigma_e$ under-massive locus is disproportionately associated with particular host types and measurement approaches.

    Taken together, these trends suggest at least two plausible, non-exclusive interpretations: (i) an astrophysical dependence of the scaling relation on morphology (e.g., differences between early-type bulges and late-type/pseudobulge systems), and/or (ii) method- and selection-dependent systematics that affect subsets of the literature (including maser-selected spirals and barred galaxies). Distinguishing these possibilities requires more homogeneous direct measurements spanning host morphologies, especially below $\sigma_e \sim 100\,\kms$. Adding direct, stellar-dynamical anchors in the $\sigma_e\lesssim 60\,\kms$ regime is therefore a key deliverable of this program.

    The morphology- and selection-dependent trends described above suggest that the apparent intermediate-$\sigma_e$ offset should not automatically be interpreted as a universal change in the scaling relation. NGC~5102 provides a direct low-$\sigma_e$ test of whether this offset persists to lower dispersions. At $\sigma_e\simsym46\,\kms$, our measured $\Mbh$ remains consistent with KH13 once intrinsic scatter is included, and therefore does not strengthen the case for a global turnover in the $\Mbh$--$\sigma_e$ relation. Given the limited number of direct measurements at $\sigma_e$ less than $100\,\kms$, this single object cannot rule out more complex low-dispersion behavior. However, it suggests that the intermediate-$\sigma_e$ under-massive locus may reflect population mix and/or measurement and selection effects that are most prominent in the intermediate-dispersion regime, rather than a universal change in slope.

    Finally, this result contributes to the low-mass black hole demographic constraints that motivate seeding and occupation-fraction studies (Section~\ref{sec:intro}). A single galaxy cannot distinguish among seed channels, but a uniform set of direct dynamical measurements across a controlled sample can provide the empirical leverage needed to test whether scaling relations remain continuous, whether intrinsic scatter increases, and how frequently low-mass galaxies host central massive black holes.

    The broader importance of this measurement is therefore methodological as well as astrophysical. By coupling a nucleus-aware stellar mass model (a variable $M/L$ tied to the NSC/bulge decomposition) with STIS kinematics that resolve the central dispersion rise and MUSE constraints that anchor the
    surrounding potential, we reduce the dominant low-mass systematic: the partition of unresolved nuclear
    mass between stars and a central point mass. Applied uniformly to more systems spanning the MBH mass spectrum, this workflow can deliver a homogeneous set of direct, stellar-dynamical black hole masses in regimes where indirect estimators are least secure, improving empirical constraints on low-mass scaling relations and on the occupation fractions relevant to seeding and early-growth models.

\section{Summary} \label{sec:summary}

    We present a stellar-dynamical measurement of the central black hole mass in the SA0$^-$ galaxy NGC~5102, improving direct constraints on black hole demographics in the under-sampled low-mass regime. Using \textit{HST}/WFPC2 imaging, \textit{HST}/STIS spectroscopy, and VLT/MUSE IFU data, we modeled the galaxy with axisymmetric three-integral Schwarzschild orbit libraries and included a radially varying stellar $M/L$ to account for the blue nuclear star cluster.
    
    Our best-fit model yields $\Mbh=(1.30^{+0.19}_{-0.18})\times10^6\,\Msun$, with $\Upsilon_V=0.48\pm0.01\,\MsunLsun$, $\xi=0.025^{+0.021}_{-0.016}$, and $i=88.5^{+0.4}_{-0.5}$ deg. A model with no central black hole is strongly disfavored ($\Delta\chi^2=90$), indicating that the nuclear LOSVDs require a compact central mass in addition to the stellar component.
    
    The variable $M/L$ treatment is essential for this system: it accommodates the blue nuclear star cluster and reduces the risk of biasing $\Mbh$ low by over-assigning stellar mass to the nucleus. The orbit solutions also indicate a multi-component internal structure, with a tangentially biased inner region and a more radially biased outer region, but the exact transition radii are not uniquely determined and should be interpreted only as a qualitative indicator of changing orbital structure.
    
    Our measurement is consistent with the prior JAM-based dynamical result for NGC~5102 \citep{Nguyen2018, Nguyen2019} at the $\sim1.7\sigma$ level (0.16 dex difference). In the $\Mbh$--$\sigma_e$ plane, NGC~5102 lies $\approxsym0.40$ dex above the KH13 relation but remains consistent within the intrinsic scatter, providing a direct low-$\sigma_e$ anchor for a regime where dynamical measurements are still sparse. The result also supports the use of high-resolution nuclear spectroscopy combined with wide-field IFU constraints and nucleus-aware stellar mass models for robust black hole mass measurements in low-mass galaxies.
    
    Extending this uniform, direct stellar-dynamical approach to a larger sample of systems will turn individual case studies into a controlled benchmark set. That benchmark will sharpen tests of low-mass scaling-relation continuity and intrinsic scatter and will provide more direct inputs for black hole occupation-fraction and seeding models that are otherwise calibrated largely by indirect inferences.

\section*{Acknowledgments}

    TKW thanks Dieu D. Nguyen for insightful conversations regarding their previous black hole mass measurement of NGC~5102. Our MUSE data were obtained from the ESO Science Archive Facility \citep{muse_eso_data}. This research was supported in part through computational resources and services provided by Advanced Research Computing at the University of Michigan. The authors acknowledge the use of U-M GPT in editing this manuscript.

\software{Astropy \citep{astropy2013,astropy2018,astropy2022},
          SAOImageDS9 \citep{Joye2003}, 
          XVISTA \citep{XVISTA},
          SciPy\_python \citep{Oliphant2007}, 
          emcee \citep{emcee}, 
          corner.py \citep{cornerpy}.}
\bibliography{references}{}
\bibliographystyle{aasjournalv7}

\end{document}